\newcommand{\half}{\frac{1}{2}}
\newcommand{\beq}{\begin{equation}}
\newcommand{\eq}{\end{equation}}
\newcommand{\bea}{\begin{eqnarray}}
\newcommand{\ea}{\end{eqnarray}}
\newcommand{\p}{\partial}
\newcommand{\nn}{\nonumber}

\documentclass[a4paper,12pt]{article}

\pdfoutput=1
\usepackage{jheppub}

\title{Intermediate scalings in holographic RG flows and conductivities}

\author[\text{$\natural$}]{Jyotirmoy Bhattacharya}
\author[\text{$\clubsuit$}]{, Sera Cremonini}
\author[\text{$\spadesuit$,$\blacklozenge$,$\bigstar$}]{and Blaise Gout\'eraux}

\affiliation[\text{$\natural$}]{Kavli Institute for the Physics and Mathematics of the Universe (WPI),
 The University of Tokyo, Kashiwa, Chiba 277-8583, Japan}
\affiliation[\text{$\clubsuit$}]{George and Cynthia Mitchell Institute for Fundamental Physics and Astronomy,
 Texas A\&M University, College Station, TX 77843--4242, USA}
\affiliation[\text{$\spadesuit$}]{Nordita, KTH Royal Institute of Technology and Stockholm University
Roslagstullsbacken 23, SE-106 91 Stockholm, Sweden}
\affiliation[\text{$\blacklozenge$}]{Stanford Institute for Theoretical Physics, Department of Physics, Stanford University,
Stanford, CA 94305, USA}
\affiliation[\text{$\bigstar$}]{APC, Universit\'e Paris 7, CNRS, CEA, Observatoire de Paris, Sorbonne Paris Cit\'e, F-75205, Paris Cedex 13, France}

\emailAdd{jyotirmoy.bhattacharya@ipmu.jp, sera@physics.tamu.edu, gouterau@stanford.edu}

\abstract{
We construct numerically finite density domain-wall solutions which interpolate between two $AdS_4$ fixed points
and exhibit an intermediate regime of hyperscaling violation, with or without Lifshitz scaling.
Such RG flows can be realized in gravitational models containing a dilatonic
scalar and a massive vector field with appropriate choices of the scalar potential and couplings.
The infrared $AdS_4$ fixed point describes a new ground state for strongly coupled quantum systems
realizing such scalings, thus avoiding
the well-known extensive zero temperature entropy associated with $AdS_2 \times \mathbb{R}^2$.
We also examine the zero temperature behavior of the optical conductivity in these backgrounds and identify two scaling regimes before the UV CFT scaling is reached.
The scaling of the conductivity is controlled by the emergent IR conformal symmetry at very low frequencies,
and by the intermediate scaling regime at higher frequencies.
}

\begin{document}

\begin{flushright}
\small{IPMU14-0297\\ MIFPA-14-27\\ NORDITA-2014-104\\SU-ITP-14/21}
\end{flushright}

\maketitle
\flushbottom
%

\section{Introduction and Summary of Results}
\label{Intro}

In recent years novel gravitational solutions have served as a holographic laboratory
for exploring phases of quantum matter with strongly coupled degrees of freedom,
which are notoriously difficult to study using conventional field theory methods.
As an example, non-relativistic scaling solutions at finite density have been used as toy models for describing some of the
unusual properties of non-Fermi liquids and strongly correlated electron systems.
In this note, we are interested in a class of gravitational solutions
which exhibit anisotropic Lifshitz scaling and realize geometrically the notion of hyperscaling violation, which were classified in \cite{Charmousis:2010zz,Gouteraux:2011ce}.
We will focus in particular on their behavior at zero temperature, thus probing the possible ground states of such systems.
While these geometries were originally thought to have features reminiscent of systems with a Fermi surface \cite{Ogawa:2011bz,Huijse:2011ef,dong},
it is now understood that such an interpretation suffers from a number of subtleties \cite{Hartnoll:2012wm,Anantua:2012nj,Edalati:2013tma}.
More broadly, they may be relevant for describing emergent infrared (IR) phases which break Lorentz invariance
while respecting rotational and translational symmetry.

Hyperscaling violating geometries with Lifshitz-like scaling are exact solutions to simple Einstein-Maxwell-dilaton (EMD) models
\cite{Charmousis:2009xr,Gubser:2009qt,Cadoni:2009xm,Charmousis:2010zz,Perlmutter:2010qu,Iizuka:2011hg,Gouteraux:2011ce}
where, however, they are often associated with curvature and tidal singularities
(see the discussions in \cite{Charmousis:2010zz} and \cite{Horowitz:2011gh,Bao:2012yt,Copsey:2012gw}).
More importantly, in such models they are supported by a running dilatonic scalar
which drives the theory to either strong or weak coupling as the IR is approached.
As a result, in general these solutions are not believed to be a good description of the geometry in the deep IR,
where they are expected to be modified\footnote{As a caveat, in a number of cases the geometry would decompactify, leading
to an alternative resolution \cite{Gouteraux:2011ce, Gouteraux:2011qh, Charmousis:2012dw, Gouteraux:2012yr}. Also, for a special class of hyperscaling violating geometries
a non-singular global extension is possible \cite{Lei:2013apa}.}.
Indeed, in a number of constructions they have been shown to arise in the intermediate region
of more complicated
spacetimes, whose deep IR `completion' is $AdS_2 \times \mathbb{R}^2$   (and sometimes a geometry conformal to it).
In particular, renormalization group (RG) flows connecting $AdS_2 \times \mathbb{R}^2$ in the deep IR to an intermediate
hyperscaling violating and Lifshitz-like geometry -- while approaching $AdS_4$ in the UV --
have been seen in \cite{Bhattacharya:2012zu,Kundu:2012jn,Donos:2012yi,Barisch-Dick:2013xga,O'Keeffe:2013nha,Ghodrati:2014spa}
(see also \cite{Harrison:2012vy,Knodel:2013fua} for the case which respects hyperscaling).

However, $AdS_2 \times \mathbb{R}^2$ is well-known to suffer from an extensive ground state entropy. 
Although it is still unclear whether it is a pathology or a feature,
such a highly degenerate ground state has sparked interest in searching for alternative IR completions.
The ground state entropy puzzle, combined with the fact that $AdS_2 \times \mathbb{R}^2$ is often unstable
to striped instabilities (see e.g. \cite{Nakamura:2009tf,Ooguri:2010kt,Donos:2011bh,Donos:2011qt,Donos:2011pn}),
has lead to the suggestion that spatially modulated phases may be a natural ground state for
\emph{certain} classes of systems with Lifshitz scaling and hyperscaling violation \cite{Cremonini:2012ir,Iizuka:2013ag,Cremonini:2013epa}.

Here we would like to focus on yet another possibility and ask whether $AdS_4$ could be a
viable candidate for a stable ground state for (non-relativistic) geometries which violate hyperscaling.
Such an $AdS_4$ IR completion would signal the presence of an \emph{emergent} conformal symmetry in the IR of the dual theory
-- analogous to that of \cite{Gubser:2009cg,Horowitz:2009ij}  -- and would also guarantee a sensible thermodynamic description\footnote{See
\cite{Amado:2014mla} for other constructions of $AdS_4 \rightarrow AdS_4$ domain wall backgrounds, and \cite{Donos:2013woa} for $AdS_5$ to $AdS_5$ superconducting p-wave domain-walls.}.
However, unlike in the setup of the ground state of the holographic superconductor \cite{Gubser:2009cg,Horowitz:2009ij},
in our construction the
intermediate regime will exhibit non-trivial scaling behavior,
including in general the breaking of Lorentz invariance.
Since we are interested in UV fixed points with full conformal invariance, we will require the geometry
to approach $AdS_4$ at the boundary\footnote{For recent constructions of flows between $AdS$ in the UV and
hyperscaling violating geometries in the IR see \cite{Dey:2014yra}.}.
We will engineer a four-dimensional model which supports finite density, zero temperature
geometries interpolating between two (distinct) $AdS_4$ solutions in the IR and UV, while traversing an intermediate region of
hyperscaling violation and Lifshitz scaling.
We will examine the conditions for the existence of such RG flows, and realize explicitly a few cases numerically.

Armed with the domain walls we have constructed, it is natural to ask whether the intermediate scaling regime
will leave an imprint on transport, and in particular on the conductive properties of the putative dual quantum system.
To this end, we will examine the zero temperature behavior of the optical conductivity and find that it exhibits
two distinct scaling regimes. The emergent IR conformal symmetry dictates the scaling of the conductivity
at very low frequencies \cite{Gubser:2009cg}. On the other hand, at intermediate frequencies
its scaling is controlled by the hyperscaling violating regime \cite{Gouteraux:2013oca}.
Finally, in the far UV the conductivity will settle to a constant, as expected for $AdS_4$ asymptotics.
The three regimes and the transition between them will be clearly visible in our numerics.
As we will see, depending on the detailed structure of the background solutions
we will be able to generate both positive and negative scalings for the conductivity in the intermediate region. The possible formation of bound states, in particular for negative intermediate scalings, may spoil this picture, so that there is a nontrivial connection between the succession of scaling regimes in the gravity background and in the optical conductivity.
Our work is a step towards realizing holographic duals to condensed matter systems which display
intermediate scalings\footnote{Note that recent results in the literature \cite{lattice1,lattice2} where such scaling
behaviour was reported in connection to inhomogenous holographic lattices are the subject of some debate \cite{Donos:2013eha, Donos:2014yya}.
In \S\ref{ssec:numcond}, through a convenient choice of parameters, we will exhibit
an intermediate scaling in the frequency dependence of the optical conductivity of the form $\text{Re}(\sigma) \sim \omega^{-2/3}$.
Unlike \cite{lattice1,lattice2}, in our case the reason for this behaviour is simply an intermediate scaling background,
with hyperscaling violation.}
in optical conductivities \cite{vandermarel1,vandermarel2}.

The types of RG flows we have here, linking a regime of hyperscaling violation at some energy scale to
a $2+1$ conformal field theory at lower energies,
may also be relevant to condensed matter systems which exhibit dimensional crossover --
as well as to systems with a Fermi surface, as mentioned above.
Dimensional crossover refers to $D$ dimensional systems which are formed by stacking together $D-1$ dimensional manifolds.
Consider the case in which the coupling inside each manifold is much stronger than that
between different manifolds.
Then at low energies (much smaller than the energy scale of intra- and inter-manifold couplings)
the system shows $D$ dimensional behavior with no violation of hyperscaling.
However, when the energy is higher than the inter-manifold couplings but smaller than the intra-manifold couplings,
the system behaves like a number of decoupled $D-1$ dimensional systems, i.e. exhibits hyperscaling violation with $\theta=1$.
An example of this behavior occurs in certain classes of
elastic models in which there are two phonon modes (two polarizations).
At high energy, the dispersion relation of each mode only depends on one of the two momenta (either $k_x$ or $k_y$)
and the system behaves like a $1+1$ dimensional system, violating hyperscaling.
On the other hand, at low energy the phonon modes are described by a $2+1$ CFT.

The key physics of this particular elastic model -- dimensional crossover, as described above -- can arise in a wide range of systems,
including strongly correlated ones.
For an example relevant to systems with a Fermi surface, we can mention the case of superconductors with nodes, such as a d-wave superconductor.
At low energy the nodal quasi particle is described by a massless Dirac theory
(a $2+1$ CFT for a superconductor in $D=2+1$ dimensions).
At high energy (much larger than the superconductor gap), the system behaves like a Fermi liquid (with hyperscaling violation $\theta=D-1$).
In both classes of systems we see a flow that is \emph{qualitatively} similar to the one we have constructed, from a
higher energy regime of hyperscaling violation to a low-energy description in terms of a $2+1$ CFT.
Clearly, it would be interesting to make more quantitative connections to real systems such as these\footnote{We are grateful to
Kai Sun for pointing these out to us.}.

The structure of the paper is as follows.
We introduce our model in Section \ref{Section2}, and discuss the IR and UV perturbations needed to trigger RG flow in Section \ref{AdS4fixedpoint}.
Section \ref{PotentialChoices} contains the choice of scalar couplings and potential needed
to obtain the intermediate scaling regime, the form of the solutions we expect to find there, and our numerical RG flows.
The frequency dependence of the conductivity is analyzed in Section \ref{Conductivity}. We conclude in Section \ref{Outlook} with
a discussion of implications and future directions.

\section{The Setup}
\label{Section2}

We are interested in working with a neutral scalar coupled to a massive vector field
\beq
\label{LagMassive1}
\mathcal{L} = R - \half (\partial \phi)^2 - \frac{1}{4} Z(\phi) \, F^2   - \half W(\phi) A^2  - V(\phi) \, ,
\eq
with the couplings $Z(\phi), W(\phi)$ and the potential $V(\phi)$ for now left completely arbitrary.
We will specify what form they should take to support the $AdS_4 \rightarrow AdS_4$
flow we are after -- and in particular the intermediate scaling regime -- in Section \ref{PotentialChoices}.
Although the model  (\ref{LagMassive1}) is completely phenomenological at this stage,
for appropriate choices of couplings 
it can be interpreted as providing an effective description of the broken-symmetry phase of a theory with
a $U(1)$ symmetry and a charged complex scalar.
With this motivation in mind\footnote{The St\"uckelberg mechanism would be an alternative way
to restore gauge invariance.},
and to be consistent with what one would obtain from the condensation of a charged scalar,
we will choose $W(\phi)$ so that it vanishes in the UV
and behaves as $\sim \phi^2$ slightly away from it.
We expect the two theories to lead to qualitatively similar physics, and
choose to adopt (\ref{LagMassive1})
to make direct contact with the analysis of \cite{Gouteraux:2012yr}, where scaling solutions to this model were studied extensively.
See also \cite{Iizuka:2012pn,Gath:2012pg}, where similar results appeared.

The equations of motion for the system (\ref{LagMassive1}) are given by
\bea
\label{Einstein1}
&& R_{\mu\nu} + \frac{Z}{2} \, F_{\mu\rho} F^{\rho}_{\;\;\nu} -  \half \p_\mu \phi \, \p_\nu \phi - \frac{W}{2} A_\mu A_\nu +
\frac{g_{\mu\nu}}{2} \left[\half(\p \phi)^2 + V -R   +  \frac{W}{2} \, A^2 +  \frac{Z}{4}  F^2  \right] = 0 \, , \nn \\
\label{scalar0}
&& \frac{1}{\sqrt{-g}}\, \p_\mu \left(\sqrt{-g}\,  \p^\mu \phi  \right) = \frac{1}{4} \frac{\p Z}{\p \phi} \, F^2 + \half \frac{\p W}{\p \phi} A^2
+ \frac{\p V}{\p \phi}  \, , \nn \\
\label{gauge0}
&& \frac{1}{\sqrt{-g}} \, \p_\mu \left(\sqrt{-g} \, Z \, F^{\mu\nu}\right) = W A^\nu  \, .
\ea
After eliminating the Ricci scalar, the Einstein equations can be expressed in the simpler form
\beq
\label{Einstein2}
R_{\mu\nu} = \half \p_\mu \phi \, \p_\nu \phi + \frac{W}{2} A_\mu A_\nu
+\frac{Z}{2}\left[ \, F_{\mu\rho} F_{\nu}^{\;\;\rho} - \frac{g_{\mu\nu}}{4}F^2  \right] +\frac{V}{2} g_{\mu\nu}\,.
\eq
Since we are interested in purely electric solutions, we choose the gauge field to be
\beq
\label{backgroundgg}
A_\mu = \left( A_t(r), 0, 0, 0 \right)\, ,
\eq
and parametrize the rest of our ansatz\footnote{As can be seen from the form of the metric, there is still some residual gauge symmetry,
which we have left unfixed because of later convenience.} by
\bea
\label{metricansatz}
&& ds^2 = - D(r) dt^2  + B(r) dr^2  + C(r) (dx^2 + dy^2) \, , \\
&& \phi=\phi(r) \, .
\ea
Einstein's equations then take the form
\bea
\phi^{\prime \, 2} &+& \frac{D^{\prime}}{D} \left(\frac{C^{\prime}}{C}-\frac{B^{\prime}}{B} -\frac{D^{\prime}}{D} \right)
- \frac{C^{\prime}}{C} \left(\frac{B^{\prime}}{B}+\frac{C^{\prime}}{C}\right) +
2 \left( \frac{C^{\prime\prime}}{C} + \frac{D^{\prime\prime}}{D} \right) + \nn \\
& -& \frac{B}{D} \, W A_t^2  + 2B V -  \frac{Z}{D} A_t^{\prime \, 2} =0 \, , \nn \\
 \phi^{\prime \, 2} &+& \frac{B}{D} \, W A_t^2  + \frac{Z}{D} A_t^{\prime \, 2}+ 2 B V -  \frac{C^{\prime \; 2}}{C^2} + \frac{4 C^{\prime\prime}}{C}
- 2 \frac{B^\prime C^\prime}{B C} = 0 \, , \nn \\
 \phi^{\prime \, 2} &+& \frac{B}{D} \, W A_t^2 - \frac{Z}{D} A_t^{\prime \, 2} -
2 B V - \frac{C^{\prime \; 2}}{C^2}- 2 \frac{D^\prime C^\prime}{D C} = 0 \, ,
\label{Einstein3}
\ea
with primes denoting radial derivatives, $^\prime \equiv \p_r$.
Finally, the scalar field and gauge field equations of motion become, respectively,
\bea
\label{scalar}
&& \frac{1}{\sqrt{B D C^2}} \; \p_r \left( \sqrt{\frac{D C^2}{B}} \, \p_r \phi \right) + \frac{\p_\phi Z}{2 B D} (\p_r A_t)^2
+ \frac{\p_\phi W}{2 D} (A_t)^2 - \p_\phi V = 0 \, , \\
\label{gauge}
&& \p_r \left( Z \sqrt{\frac{ C^2 }{D B}}\p_r A_t \right) = W \sqrt{\frac{B C^2}{D}} A_t \, .
\ea
We now have all the ingredients we need to examine the structure of the solutions to this theory.

\section{RG Flow}
\label{AdS4fixedpoint}

We are interested in constructing domain wall solutions which flow between two $AdS_4$ fixed points, while traversing
an intermediate `scaling' regime exhibiting anisotropic Lifshitz scaling $z\neq 1$ and hyperscaling violation $\theta\neq0$.
In particular, we want the scalar field in our model to roll from a maximum of the effective potential in the UV to a minimum in the IR,
taking on constant values $\phi = \{ \phi_{UV},\phi_{IR} \}$ at the endpoints of the flow,
where the metric should be of the $AdS_4$ form
\beq
\label{ads4}
ds^2_{AdS_4} =  - r^2 \, dt^2 + L^2 \, \frac{dr^2}{r^2} + r^2 (dx^2 + dy^2)  \, .
\eq
We use $L=\{L_{IR},L_{UV}\}$ to parametrize the size of the $AdS_4$ radius in the IR and UV, respectively.
The effective scalar field potential for our model (\ref{LagMassive1})
\beq
\label{Veff}
V_{eff}(\phi) = V(\phi) + \frac{1}{4} Z(\phi) F^2 + \half W(\phi) A^2 \, ,
\eq
will therefore have to be engineered to admit two extrema,
\beq
\label{Vmin}
V_{eff}^\prime (\phi=\phi_{IR}) = 0 \quad \quad \text{and} \quad \quad V_{eff}^\prime (\phi=\phi_{UV}) = 0 \, ,
\eq
located at the IR and UV values of the scalar.
From now on we will use
\beq
\phi_{IR} = \phi_0  \, , \quad \quad \text{and} \quad \quad \phi_{UV}=0 \, ,
\eq
the UV value being chosen to be at zero purely for convenience.
We will come back to the particular choice of potential and scalar couplings needed to achieve
the intermediate $\{z,\theta\}$ scaling region in the next section.

Note that in theories of the type of (\ref{LagMassive1}), the requirement (\ref{Vmin}) that
the scalar sits at an extremum of its potential also allows  -- under appropriate conditions on the couplings --
for an IR $AdS_2 \times \mathbb{R}^2$ solution as well as Lifshitz geometries. Here we will choose $W(\phi)$ such that no $AdS_2\times \mathbb{R}^2$ can exist, though a Lifshitz solution may.
To ensure that the infrared geometry reached by our flows is indeed $AdS_4$, the current dual to the gauge field must be \emph{irrelevant}
about the IR fixed point \cite{Gubser:2009cg}.
By substituting (\ref{ads4}) and the constant scalar ansatz $\phi=\phi_{IR}$ into (\ref{Einstein2}),
it is easy to check that the equations of motion are only satisfied when the gauge field vanishes in the IR,
\beq
A_t\vert_{IR} = 0 \, .
\eq
In fact, to realize our RG flow there should not be any relevant deformations at all about the IR $AdS_4$ geometry.
The irrelevant scalar and gauge field perturbations will then be responsible for connecting the geometry to the
intermediate scaling solution and to the UV fixed point.

\subsection{IR Perturbations}
\label{Deformations}

We are now ready to examine the structure of the perturbations about the IR fixed point.
We expand around the IR $AdS_4$ solution in the following way,
\bea
\label{IRexp}
&& B(r) = L_{IR}^2 \; r^{-2} \left(1+ B_1 r^{-b_1} \right) \, , \quad
D(r) = r^{2} \left(1+ D_1 r^{-d_1} \right) \, , \quad
C(r) = r^{2} \left(1+ C_1 r^{-c_2} \right) \, , \nn \\
&& \phi(r) = \phi_0 + \phi_1 r^{-\beta^\phi} \, , \quad \quad \quad \quad A_t(r) = r^{-c_0} \left(Q + A_1 r^{-\beta^A} \right) \, ,
\ea
and furthermore assume that $Q=c_0=0$ to ensure that the gauge field vanishes to leading order.
It is then easy to see from the equations of motion (\ref{Einstein2}) that the potential at the IR fixed point is
\beq
V(\phi_0)=-\frac6{L^2_{IR}} \, .
\eq
For $AdS_4$ to be the true ground state of the system only irrelevant deformations
must be turned on there.
With our coordinate choice  the boundary is at $r\rightarrow\infty$ while the deep infrared is located at $r=0$.
Thus, a mode $\delta h $ scaling as $\delta h  \sim r^{-\beta}$ can only be irrelevant provided that  $\beta < 0$.
Note that -- since in the IR the geometry is supported by a constant scalar and vanishing charge --
to linear order the metric fluctuations decouple from those of the scalar and gauge fields.
To higher order in perturbations, however, the couplings between all modes will have to be taken into account,
which will be done in the numerics.
Here we work to linear order and focus entirely\footnote{The remaining deformations of the metric
correspond to a constant shift in time or to turning on a non-zero temperature, taking us away from extremality.} on deformations
which are triggered by $\delta\phi$ and $\delta A_t$:
\begin{itemize}
\item
\underline{scalar perturbations}\\ \\
Taking $\delta \phi = \phi_1 r^{-\beta^\phi}\neq 0$, one finds that the scaling exponent $\beta^\phi$ must obey
\beq
(\beta^\phi)^2 - 3 \beta^\phi - L_{IR}^2 V^{\prime\prime}(\phi_{0}) = 0 \, , 
\eq
with solutions given by
\beq
\beta^\phi{_\pm} = \frac{3}{2} \pm \sqrt{ \frac{9}{4} + L_{IR}^2 \, V^{\prime\prime}(\phi_{0}) }
=\frac{3}{2} \pm \sqrt{ \frac{9}{4} + L_{IR}^2 \, m_{IR}^2 } \; ,
\eq
where we have traded the potential for the mass of the scalar, $V^{\prime\prime}(\phi_{0}) = m_{IR}^2$.
To avoid complex scaling dimensions and hence potential IR instabilities we assume first
\beq
9 + 4 L_{IR}^2 \, m_{IR}^2 >0 \, ,
\eq
which is nothing but the $AdS_4$ BF bound condition.
Moreover, here we will impose the stronger constraint $m_{IR}^2 >0$,
which will allow us to identify the root $\beta_-^\phi$ with the irrelevant mode.
Introducing the scaling dimension $\Delta_\phi$ of the operator ${\cal O}_{IR}$
dual to the scalar $\phi$,
\beq
\label{Deltaphi}
\Delta_\phi = \frac{3}{2} + \sqrt{\frac{9}{4}+m_{IR}^2 L^2_{IR}} \, ,
\eq
the fluctuation $\delta\phi\sim r^{-\beta_-^\phi}$ can then be rewritten as
\beq
\label{irrscalar}
\delta\phi \sim r^{\Delta_\phi-3} \, .
\eq
The statement that the scalar field mode is irrelevant can then be phrased as
\beq
\boxed{\Delta_\phi = \frac{3}{2} + \sqrt{\frac{9}{4}+m_{IR}^2 L^2_{IR}}  >3 \, . \;}
\eq

\item
\underline{gauge field perturbations}\\ \\
Looking now at the equation of motion for $\delta A_t = A_1 r^{-\beta^A} \neq 0$, we find
\beq
+ 6 W(\phi_0) - \beta^A V(\phi_0) Z(\phi_0) + (\beta^A)^2 V(\phi_0) Z(\phi_0) = 0 \, , \quad \quad 24 W(\phi_0) - V(\phi_0) Z(\phi_0) >0 \, ,
\eq
with solutions
\beq
\beta^A_\pm = \half \pm \half \sqrt{1+\frac{4L_{IR}^2 \, W(\phi_{0})}{Z(\phi_{0})}} \, .
\eq
Notice that the choice $W(\phi_{0}) =0$ corresponds to $\beta^A = \{0,1\}$, hence the gauge field perturbations
are either marginal or relevant.
This reiterates the need for a non-zero gauge field mass $W \neq 0$ if we want
$AdS_4$ to describe the IR fixed point of the theory.\footnote{However, a hyperscaling violating fixed point with $z=1$ could be
reached with a massless gauge field $W(\phi)=0$, i.e. the non-zero $\theta$ can decrease appropriately the conformal dimensions
of the dual conserved current \cite{Gouteraux:2012yr}.}
As mentioned above for the scalar field case,
we are not interested in triggering any infrared instabilities, and will therefore choose parameters to avoid
any tachyonic masses -- we will restrict the scaling dimensions to be real.

Then, under the assumption that $W(\phi_{0})$ and  $Z(\phi_{0})$ are both positive,
the perturbation which vanishes in the deep IR (as $r \rightarrow 0$) is the one corresponding to the $\beta_-^A$ root.
In the gauge field case, however, the discussion of which modes trigger irrelevant deformations is more subtle.
Introducing the conformal dimension $\Delta_J$ of the operator
$J_0^{IR}$ dual to the gauge field $A_t$,
\beq
\label{DeltaJ}
\Delta_J = \frac{3}{2} + \half \sqrt{1+ \frac{4 L_{IR}^2 \, W(\phi_0)}{Z(\phi_0)}} \, ,
\eq
we can rewrite the gauge field perturbation as
\beq
\label{ggpert}
\delta A^{-}_t \propto r^{\Delta_J-2} \, .
\eq
As discussed in \cite{Gubser:2009cg}, the gauge field is dual to an \emph{irrelevant current} when
$\Delta_J >3$, i.e. when
\beq
\boxed{\; \sqrt{1+ \frac{4 L_{IR}^2 \, W(\phi_0)}{Z(\phi_0)}} > 3 \, . \; }
\eq
This is the case which we are interested in, which will ensure that the field theory current
will drive the flow towards Lorentz-invariant conformality in the IR.

Note that even though for $ 2<\Delta <3$ the gauge field perturbation vanishes as $r\rightarrow 0$,
 the dual field theory current is \emph{relevant} in that range. Inspecting the backreaction on the metric fields at
 quadratic order, we find modes like $r^{-2(\beta_a-1)}=r^{2(\Delta_J-3)}$.
 They will destroy the IR $AdS_4$ and other scaling solutions are expected to arise in the deep infrared. If $W(\phi_{IR})\neq0$,
 the flow will be driven to a Lifshitz fixed point, while if $W(\phi_{IR})=0$ it will land in an $AdS_2\times \mathbb{R}^2$
 (we refer the reader to \cite{Gubser:2009cg, Gouteraux:2012yr} for a more detailed discussion).

\end{itemize}

\subsection{UV Restrictions}

In order to generate RG flow away from the UV fixed point, we would like to deform
the conformal field theory dual to the UV $AdS_4$ solution by turning on relevant
deformations of the scalar and gauge field.
In the UV (as $r\rightarrow 0$ in our notation), the scalar has a boundary expansion of the form
\beq\label{UVsclform}
\phi\sim\phi_{UV}+\frac{\Phi_1}{r^{3-\Delta}}+ \frac{\Phi_2}{r^{\Delta}} +\dots
\eq
where for us $\phi_{UV}=0$. The scaling dimension $\Delta$ is fixed by the scalar potential
\beq\label{UVsd}
\Delta=\frac32+\sqrt{\frac94+m^2_{UV}L^2_{UV}} \, ,
\eq
with
\beq
V(\phi_{UV})=-\frac6{L^2_{UV}}\,,\qquad V'(\phi_{UV})=0\,,\qquad  V''(\phi_{UV})=m^2_{UV} \, .
\eq
For the usual quantization and appropriate values of $m^2_{UV}$, $\Phi_1$ corresponds to the source of
the relevant scalar operator dual to $\phi$ with conformal dimension $\Delta<3$,
as needed to trigger RG flow away from the UV fixed point.
Here we will not insist that the breaking of the $U(1)$ symmetry is spontaneous -- although the latter can in principle
be engineered.

We would also like the gauge field to have the standard asymptotic expansion, of the form
\beq
A_t=\mu+ \frac{\rho}{r} +\dots \, ,
\eq
with $\mu$ the chemical potential and $\rho$ the charge density.
This fixes the conformal dimension of the dual current to be $\Delta=2$, which again is relevant.
As can be seen by inspecting the gauge field equation of motion (\ref{gauge}), this puts constraints on the UV value of the coupling $W$, namely
\beq
\boxed{W(\phi_{UV})=0 \, . \; }
\label{WUVcond}
\eq

\section{Choice of Couplings and Potential}
\label{PotentialChoices}

In Section \ref{AdS4fixedpoint} we examined the structure of the potential and the
spectrum of perturbations about the two $AdS_4$ fixed points, setting up the conditions for the RG flow.
We are now ready to discuss when we can expect to encounter an intermediate regime of hyperscaling violation
and anisotropic Lifzhitz scaling.
To achieve the latter, the scalar potential and gauge couplings have to be engineered appropriately.
Recall that in Einstein-Maxwell-dilaton theories of the form
\beq
\label{EMD}
\mathcal{L} = R - \half (\partial \phi)^2 - \frac{1}{4} e^{\alpha\phi} \, F^2  - V_0 e^{-\eta\phi} \, ,
\eq
the \emph{minimal} couplings needed to generate $\{z,\theta\}$ scaling solutions are
\beq
\label{HypViolScalCouplings}
Z(\phi) \sim e^{\,\alpha \phi} \, ,   \qquad V(\phi) \sim e^{-\eta\phi} \, ,
\eq
with the two lagrangian parameters $\{\alpha,\eta\}$ determining the values of the scaling exponents $\{z,\theta\}$.
However, in the simple EMD model (\ref{EMD}) the gauge field is massless and the $U(1)$ gauge symmetry is unbroken.
Moreover, the electric flux of the resulting geometries is constant throughout the bulk --
in the language of \cite{Hartnoll:2011pp}, the corresponding solutions are \emph{fractionalized}.

\begin{figure}[h!]
\centering
\includegraphics[scale=0.6]{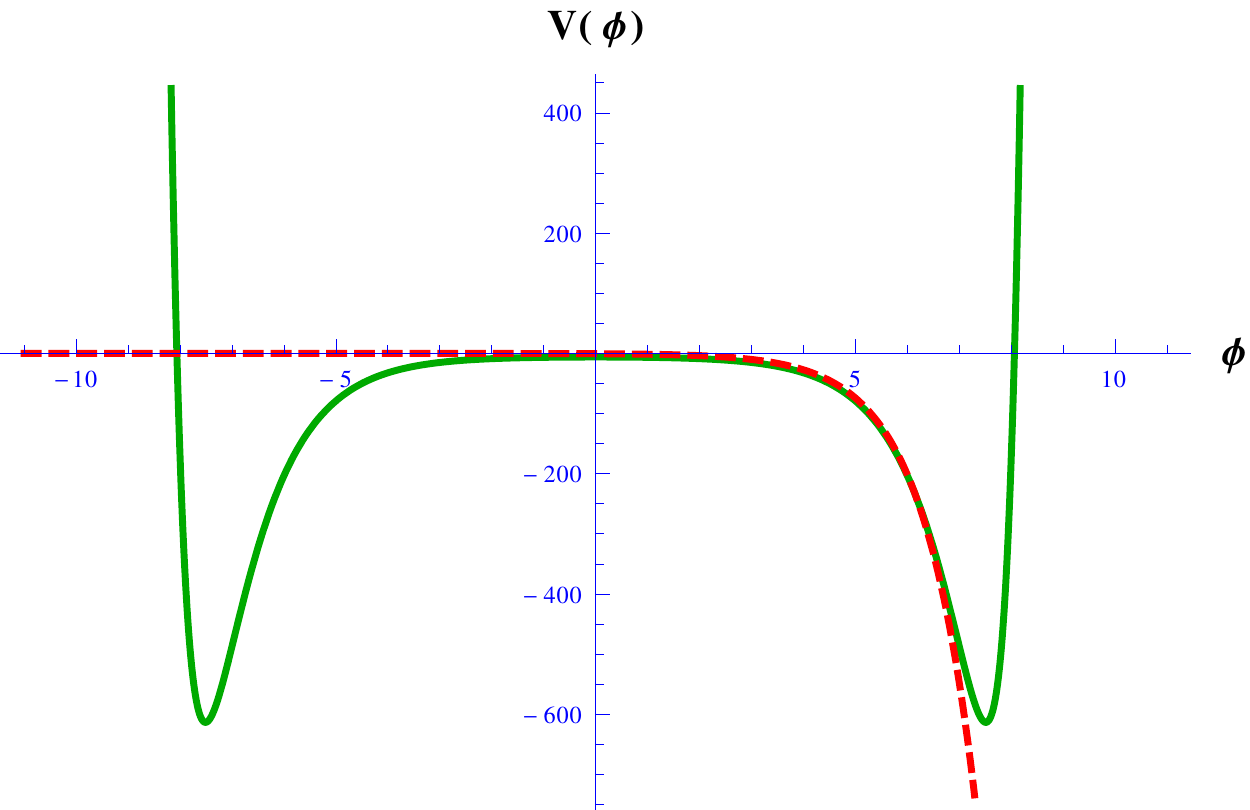}
\caption{\label{fig:scalarpot} The solid green curve is a generic plot
of the scalar potential in \eqref{Choices} which we use to construct our solutions.
The dashed red curve is a single exponential function. Note that the red curve coincides with the green
curve in a region between the maximum and the $\phi >0$ minimum. This region of overlap improves
with increase in separation between the extrema. For the solutions presented in
this paper the scalar field flows from maximum in the UV to the minimum
in the IR given by $\phi >0$.}
\end{figure}

Clearly, this is not the setup we are working with, since our infrared $AdS_4$
demands a mass term for the gauge field and places us in the symmetry-broken phase of the theory.
Furthermore, it requires that the flux should vanish at the horizon -- indeed,
in our construction the term $\sim W(\phi) A^2$ plays the crucial role of pushing the charge density
higher up in the bulk and leading to a solution which is \emph{cohesive} (borrowing the terminology of \cite{Hartnoll:2012ux}).
Taking into account the fact that the gauge couplings $Z(\phi), W(\phi)$ and scalar potential $V(\phi)$ need to satisfy (\ref{Vmin})
and to effectively reduce to (\ref{HypViolScalCouplings}) in some intermediate part of the geometry,
we choose
\beq
\label{Choices}
\boxed{
Z(\phi) =  Z_0 e^{\,\alpha\phi}  ,  \, W (\phi) =4W_0\sinh^2\left(\frac\beta2\phi\right)  ,  \,
V (\phi) = 2 V_0 \cosh\delta\phi + 2 V_1 \cosh\gamma\phi + V_3 \, . \; }
\eq
The parameters in the scalar potential $V(\phi)$ are chosen so as to have a double-well structure,
as shown by the green curve in fig. \ref{fig:scalarpot}.
The extrema of the potential correspond to distinct $AdS_4$ solutions, characterized by different radii (the two minima
correspond to the $AdS$ with smaller radii while the maximum to the one
with larger radius). Since we intend to flow from one $AdS$ to another, the holographic
c-theorem \cite{Freedman:1999gp} implies that the dilaton must flow from the maximum
in the UV to a minimum in the IR\footnote{The construction of \cite{Freedman:1999gp} can be applied to the 
relativistic domain-wall solutions we construct 
in this paper, which have an intermediate regime of hyperscaling violation. 
However, it does not apply to the more general solutions we construct, for which 
$z\neq1$ along the flow between the two conformal fixed points.
For a discussion of the breakdown of holographic
c-theorems when Lorentz invariance is broken, for geometries with an intermediate hyperscaling violating regime,
see e.g. \cite{Cremonini:2013ipa}.}. The parameters in the first two terms of $V(\phi)$
are chosen (see \S \ref{ssec:sclpotpara}) to ensure the double-well structure, while
the constant term $V_3$ is used to ensure that the UV $AdS$ radius is unity.

The crucial feature of the choice \eqref{Choices} is the fact
that, provided the minima are well separated from the maximum\footnote{For a given choice of $\gamma > \delta$,
this separation is essentially controlled by the smallness of the
parameter $V_1$ compared to $V_0$.},
there exists an intermediate region where the scalar potential is well
approximated by a single exponential function (like the dashed red curve
in fig. \ref{fig:scalarpot}). It is in this region where we may expect another
scaling solution to arise. In order for this to happen, we should also ensure that
functions $Z$ and $W$ are well approximated by single exponential functions,
when the gauge field is non-trivial.
The precise form of the resulting scaling solutions will also be sensitive to the
strength of the gauge field terms,
as well as to the interplay between $ W A^2$ and $ Z F^2$.
In general, we expect different scenarios and intermediate phases depending
on the profile of $W(\phi)$, and in particular
on where in the bulk the charge density will be concentrated.

\subsection{Constraining the Scalar Potential} \label{ssec:sclpotpara}

We would like to reduce the number of free parameters in our model (\ref{Choices}) as much as possible.
First, to make the calculations particularly tractable from now on we will take $$\gamma= 3\delta \, . $$
To ensure that the endpoints of the flow are critical points of the potential we require\footnote{Recall that
the gauge field vanishes in the deep IR.
Similarly, at the UV fixed point the contributions from the gauge field terms are subdominant
compared to those of the potential. As a result, $V_{eff}^\prime = V^\prime$ at both fixed points.}
\beq
\label{extrema}
\left.\frac{\partial V}{\partial\phi}\right|_{\phi=0}=0 \, , \qquad \qquad
\left.\frac{\partial V}{\partial\phi}\right|_{\phi=\phi_0}=0 \, .
\eq
It is immediately obvious that the UV value $\phi=0$ is an extremum of the scalar potential\footnote{Then
the leading term in the UV expansion of $W(\phi)$ is quadratic,
as in the original holographic superconductor setup \cite{HoloSc,HoloSC2}.}.
The requirement that it is a \emph{maximum} restricts the range of $V_0$ and $V_1$,
\beq
\label{modmax}
V_0 + 9 V_1 < 0 \, .
\eq
Moreover, after expanding around the UV fixed point $\phi=0$ the potential must reduce to
$$ V(\phi \sim 0) \approx - \frac{6}{L^2_{UV}} + \half \, m^2_{UV} \, \phi^2 \, . $$
Reading off the UV values of the $AdS_4$ radius $L_{UV}$ and of the scalar mass $m_{UV}$ we find
\beq \label{UVmass}
L^2_{UV} = - \frac{6}{2(V_0+V_1)+V_3} \, , \qquad m_{UV}^2 = 2\delta^2 (V_0 + 9 V_1) \, .
\eq

Let us now turn to the location of the minima of the potential, from which we will read off the IR fixed point.
When $\gamma =3 \delta$ the potential has additional critical points located at
\beq
\phi_{crit} = \frac{1}{\delta} \ln \left(\frac{\pm \sqrt{-V_1 V_0 - 3 V_1^2 \pm
\sqrt{V_0^2 + 6 V_0 V_1 - 27 V_1^2}} }{\sqrt{6} \, V_1}\right).
\eq
We will identify $\phi_0 \equiv \phi_{IR}$ with the positive (real) root which is closest to $\phi_{UV}=0$,
and which yields a minimum of the potential.
Finally, expanding the scalar potential about $\phi=\phi_0$,
$$V(\phi \sim \phi_0) \approx - \frac{6}{L^2_{IR}} + \half m^2_{IR} (\phi-\phi_0)^2 \, , $$
we can read off
\beq
L^2_{IR} = - \frac{6}{V_3 + 2 V_0 \cosh\delta\phi_0 + 2 V_1 \cosh 3\delta\phi_0} \, ,
\eq
as well as
\beq
m^2_{IR} = 2 \delta^2 \left(V_0 \cosh\delta\phi_0 + 9 V_1 \cosh3\delta\phi_0 \right) \, .
\eq
Combining the last two expressions will then give us the conformal dimension of the scalar field (\ref{Deltaphi}) in the deep IR.

\subsection{Intermediate Scaling Regime and Solutions}\label{ssec:IMHVL}

In order for the RG flow to traverse an intermediate regime of hyperscaling violation and Lifshitz scaling we would like
our model to be very well approximated by
\beq
\label{Lint}
\mathcal{L}_{int} \sim R -\half \p\phi^2 - \frac{Z_0}{4} \, e^{\alpha\phi} F^2 - \frac{W_0}{2} \, e^{\beta\phi} A^2 - V_0 e^{-\eta\phi} \, .
\eq
in an appropriate `middle' region.
This entails isolating a part of the geometry in which
the scalar potential of (\ref{Choices}) is entirely dominated by a single exponential term.
This can be achieved for example by arranging for $V_1 \ll V_0,V_3$
which suppresses the coefficient of the $\cosh 3\delta\phi$ term in the scalar potential.
The `size' of the intermediate scaling regime will become more stretched as $V_1$ grows smaller.
The intermediate scaling solutions will then be effectively controlled by the potential term
\beq
V \sim  V_0 e^{\delta\phi} \, .
\eq
Similarly, the gauge field mass term will be dominated by a single exponential,
\beq
W = 4 W_0 \sinh^2\left( \frac{\beta\phi}{2}\right) \sim  W_0 \, e^{\beta \phi} + \ldots
\eq
and we expect different scaling behaviors depending on the values of the parameters $\alpha, \beta$ and $\eta=-\delta$ as well
as of the ratio $W_0/V_0Z_0$. The structure of the geometry will also be sensitive to where the charge density will be most important.

Scaling solutions arising in the class of theories (\ref{Lint}) have been studied in detail in
\cite{Gouteraux:2012yr,Gouteraux:2013oca,Iizuka:2012pn,Gath:2012pg}
\footnote{Some algorithms for constructing the holographic dictionary with UV hyperscaling (violating) asymptotics in Einstein-Proca-Dilaton theories were recently developed in
\cite{Chemissany:2014xpa,Chemissany:2014xsa, Hartong:2014oma}. See also \cite{Gouteraux:2011qh, Charmousis:2012dw} for prior work on developing a holographic dictionary for hyperscaling violating theories via generalized dimensional reduction.}.
In particular, cohesive Lifshitz and hyperscaling violating geometries of the type we are after have been identified
when the Lagrangian parameters obey $\beta = \alpha- \eta$. Here we refer the reader to the analysis of
\cite{Gouteraux:2012yr,Gouteraux:2013oca,Iizuka:2012pn,Gath:2012pg} and simply
list the solutions which will be relevant to our discussion, following the notation of \cite{Gouteraux:2013oca}.
For appropriate parameter choices
we expect these $\{z,\theta\}$ geometries to describe the intermediate regime of our domain wall
solutions, as the latter flow between the UV and IR $AdS_4$ fixed points.
In the next subsection we will construct numerically explicit RG flows corresponding to some of these cases.
\vspace{0.2in}

\noindent
\underline{{\bf Case (i): $\beta=\alpha-\eta$, $z\neq1$}}

\vspace{0.15in}
\noindent
 If $\beta=\alpha-\eta$ and $W_0/V_0Z_0$ is determined as a function of $\alpha$, $\eta$ and $z$,
 then there is a \emph{cohesive} hyperscaling violating solution described by
 (see section 3.3.2 of \cite{Gouteraux:2012yr} for further details or \cite{Iizuka:2012pn,Gath:2012pg})
 \beq\label{EMDScalingCohesive}
	\begin{split}
		&d s^2 = -r^{2\frac{\theta-2z}{\theta-2}}d t^2+r^{\frac{4}{\theta-2}}L^2d r^2+ r^2\left(d x^2+d y^2\right),\\
		&L^2 =-4\frac{(1+z-\theta)(2+z-\theta)+(z-1)\xi}{(\theta-2)^2V_0}\,,\\
		&e^\phi=r^{\frac{2\theta}{\eta(\theta-2)}}\,, \qquad \eta = \frac{\theta }{\sqrt{2 (1-z) (\zeta-\xi) +(\theta-2 ) \theta }},\\
		& A=\sqrt{\frac{2 (z-1)}{Z_0 (z-\zeta+\xi)}}\, r^{2\frac{\zeta-\xi-z}{\theta-2}}d t\,,\qquad \alpha =\frac{\theta-2\zeta+2\xi }{\sqrt{2 (1-z)( \zeta-\xi) +(\theta-2 ) \theta }} \, ,\\
		&\frac{W_0}{Z_0}=\frac{4(z-\zeta +\xi) \xi}{(\theta-2)^2L^2}\,, \qquad \beta = \frac{-2\zeta +2\xi}{\sqrt{2 (1-z) (\zeta-\xi) +(\theta-2 ) \theta }}\,,\\
&\int e^{\alpha\phi}\star F\sim r^{2\frac{\xi}{\theta-2}},\qquad \zeta=\theta-2 \, .
	\end{split}
\eq
Notice that it is parameterized by four scaling exponents: the dynamical critical exponent $z$,
the hyperscaling violation exponent $\theta$, the cohesive (electric flux) exponent $\xi$
and the `conduction' or vector hyperscaling violating exponent $\zeta$ (which for this class of solutions is not independent\footnote{The
need for three independent scaling exponents to fully parameterize these solutions was also pointed out in \cite{Gath:2012pg}.
The parametrization used there highlighted the breaking of scale invariance by the Maxwell term in the action and differs from the one used in \eqref{EMDScalingCohesive}.}
from $\theta$). One may solve for these scaling exponents in terms of action parameters $\eta$, $\alpha$ and $W_0$. Then, as the equation relating $W_0$ is quadratic in $\xi$, we will actually obtain two solutions.

While $\zeta$ and $\xi$ do not enter in the scaling of metric components,
they do affect the particular expressions for the parameters $\alpha$ and $\eta$, as well as the radial dependence of the flux, or the values of numerical prefactors. For generic $\xi$, the electric flux vanishes in the IR of the solution $r\to 0$, as needed in our construction.
However, when $\xi=0$, the electric flux becomes constant and
the usual hyperscaling violating solutions \cite{Charmousis:2010zz} are recovered.

To ensure that this $\{z,\theta\}$ solution can be reached from the UV, it should have two irrelevant deformations.
For a detailed analysis of linear perturbations to (\ref{EMDScalingCohesive}) we refer the reader to \cite{Gouteraux:2012yr} (Appendix D.6).
In the spectrum of perturbations one finds two pairs of degenerate modes, one pair which is marginal
(describing time translations and constant shifts of the scalar) and the other one
corresponding to turning on either a finite temperature or a VEV for the scalar. Upon including subleading corrections in the scalar potential and couplings, one of the marginal modes will become irrelevant.
There are also two modes $\beta_\pm$ which maintain zero temperature,
\beq
\begin{split}
&\beta_\pm=\frac1{\theta-2}(2+z-\theta)\pm\frac1{\theta-2}\sqrt{\frac{X_m}{2(1-z)(\theta-2 -\xi )+\theta (\theta -2)}}\\
&X_m=(2-\theta ) \left(16 z^3-34-32 z^2 (\theta-1 )+47 \theta -16 \theta ^2+2 z \left(8 \theta ^2-7-8 \theta \right)\right)\\
&\quad+2 (z-1) \xi  \left(81+72 z+8 z^2-96 \theta -36 z \theta +28 \theta ^2-64 \xi -24 z \xi +32 \theta  \xi -16 \xi ^2\right).
\end{split}
\eq
To match onto the UV boundary conditions we can use one of the marginal modes
(e.g. the perturbation corresponding to the constant shift of the scalar field),
but the irrelevant mode will need to come from the $\beta_-<0$ perturbation, as $\beta_+>0$ always. This restricts the allowed parameter space.
Finally, we should require the heat capacity $\sim \frac{\partial S}{\partial T}$ to be positive, with $S\sim T^{\frac{2-\theta}{z}}$, so we will need to ensure that the scaling exponents obey
 $\frac{2-\theta}{z}>0$.
\vspace{0.2in}

\noindent
\underline{{\bf Case (ii): $\beta=\alpha-\eta$, $z=1$}}

\vspace{0.15in}
\noindent
The previous solution \eqref{EMDScalingCohesive} is ill defined for $z=1$.
Indeed, solving the field equations more carefully one may also find solutions for which $z=1$,
but in this case the electric potential decouples from the other fields (metric and scalar) and does not backreact
at leading order on Einstein's equations
or the scalar equation  (see section 3.1.2 of \cite{Gouteraux:2012yr} for more details).
In this case the metric and scalar field take the simple form
 \beq\label{EMDScalingCohesivez=1}
	\begin{split}
		&d s^2 = r^{\frac{4}{\theta-2}}\, L^2 dr^2+ r^2 \left(-d t^2+d x^2+d y^2\right),\\
		&L^2 =-\frac{4(\theta-2)(\theta-3)}{(\theta-2)^2V_0}\,,\\
		&e^\phi=r^{\frac{2\theta}{\eta(\theta-2)}}\,, \qquad \eta = \frac{\theta }{\sqrt{(\theta-2 ) \theta }} \, .
	\end{split}
\eq
For a detailed analysis of linear perturbations to (\ref{EMDScalingCohesivez=1}) we refer the reader to \cite{Gouteraux:2012yr}.
The static fluctuations around the metric (scaling schematically as $\sim r^{-\beta}$)
again appear as pairs, with a doubly-degenerate universal pair
\beq
\beta_0=0\,,\qquad \beta_u=2\frac{3-\theta}{\theta-2} \, ,
\eq
where $\beta_u$ introduces non-zero temperature and should be relevant, $\beta_u>0$.

The gauge field can simply be solved from Maxwell's equation, and there is no constraint placed on $Q$ or $W_0$:
one is simply a free integration constant, while the other appears as a parameter of the theory.
The gauge field displays two modes
\beq\label{probeAexpo}
A=Q^\pm \, r^{-\beta^q_\pm},\qquad \beta^q_+=\frac{2\xi}{\theta-2},\qquad \beta^q_-=-2\frac{\zeta-\xi-1}{\theta-2} \, ,
\eq
which backreact on the metric at quadratic order as
\beq
\beta_+=\frac{4-\theta-\zeta}{\theta-2}\,,\qquad \beta_-=\frac{\zeta-\theta+2}{\theta-2} \, .
\eq
The exponent $\xi$ is still defined from the scaling of the electric flux,
\beq
\int e^{\alpha\phi}\star F\sim r^{\frac{2\xi}{\theta-2}}\,.
\eq
For that solution to be reachable from the UV we require $\beta_-<0$,
as $\beta_+>0$ from other constraints.
Unlike in the case of the $z\neq1$ solutions, the exponent $\zeta$ is now independent from $\theta$,
and through $\beta_-$ it parametrizes the scaling dimension of the charge density in the intermediate region, which is why it can be thought of as a vector hyperscaling violating exponent.
Note that in the previous class of solutions (\ref{EMDScalingCohesive}), $\beta_-=0$ and the charge density dimension is marginal \cite{Gouteraux:2013oca}
(see also the discussion in \cite{Karch:2014mba}).
\\

Finally, note that when the gauge field is completely absent (\ref{EMDScalingCohesivez=1}) is still a valid solution.
As we will see below, even in this case our domain wall solutions will hit
(\ref{EMDScalingCohesivez=1}) while traversing the intermediate hyperscaling violating regime.

\subsection{Numerical Solutions}\label{ssec:numsol}
%
In this subsection we shall present the construction of numerical solutions
which interpolate between two $AdS_4$ geometries passing through
a hyperscaling violating (Lifshitz) region.
The existence of these different kinds of geometries, in a single domain-wall solution, will be made manifest
by exhibiting the distinctive scaling properties of the metric functions
and gauge field in the numerical plots.

For the numerics we shall make the residual gauge choice in \eqref{metricansatz} by letting
\begin{equation}\label{newgauge}
 B(r) = Y(r), \quad C(r) = X(r), \quad D(r) = \frac{1}{X(r)}.
\end{equation}
Note that this is slightly different from that used in \S\ref{ssec:IMHVL}; the
radial coordinate in \S\ref{ssec:IMHVL} is related to that in this subsection
through the relation
\begin{equation}
 r \rightarrow r^{\frac{\theta -2}{2(\theta-1)}} \, .
\end{equation}
Even in these coordinates $r\rightarrow \infty$ corresponds to the UV while $r \rightarrow 0$ to the IR.

We choose convenient values of the lagrangian parameters for obtaining the numerical solutions.
First we perform a series solution about the IR $AdS$, up to sufficiently high orders. Then we
use this series solution (with suitable values for the amplitudes of the fluctuations)
to provide boundary conditions for the numerical evolution towards the UV, making sure
we pass through an intermediate scaling regime. Finally, we use the scaling exponents
in the intermediate region to confirm the presence of a hyperscaling violating
(Lifshitz) geometry there.

Can analytical estimates of the transition be computed? For instance, one may try to compare the magnitude of the vector kinetic term with the UV/IR value of the scalar potential, as in \cite{Harrison:2012vy}, and the estimates obtained this way seem to have the right order of magnitude. However, note that these transitions are very nonlinear effects. Indeed, it can be checked by comparing the numerics to the approximate UV/IR expansions that perturbation theory completely breaks down at these scales, and thus any analytical estimates of the exit from the AdS$_4$ regimes at both ends of the flow are unreliable.

\subsubsection{Zero density flows with \texorpdfstring{$\theta \neq 0$}{theta different from 0}}
\label{sssec:onlyscalar}
%
Let us first discuss the simplest case in which we set the gauge field to zero. In this
Einstein-dilaton system we cannot obtain a Lifshitz-like scaling, i.e. we must have $z=1$. Although the
dynamical exponent is trivial throughout the flow, we can have hyperscaling violating intermediate
regimes where the solution is well approximated by \eqref{EMDScalingCohesivez=1}
(after the appropriate gauge change specified by \eqref{newgauge}).

\begin{figure}[h!]
\centering
\includegraphics[scale=0.56]{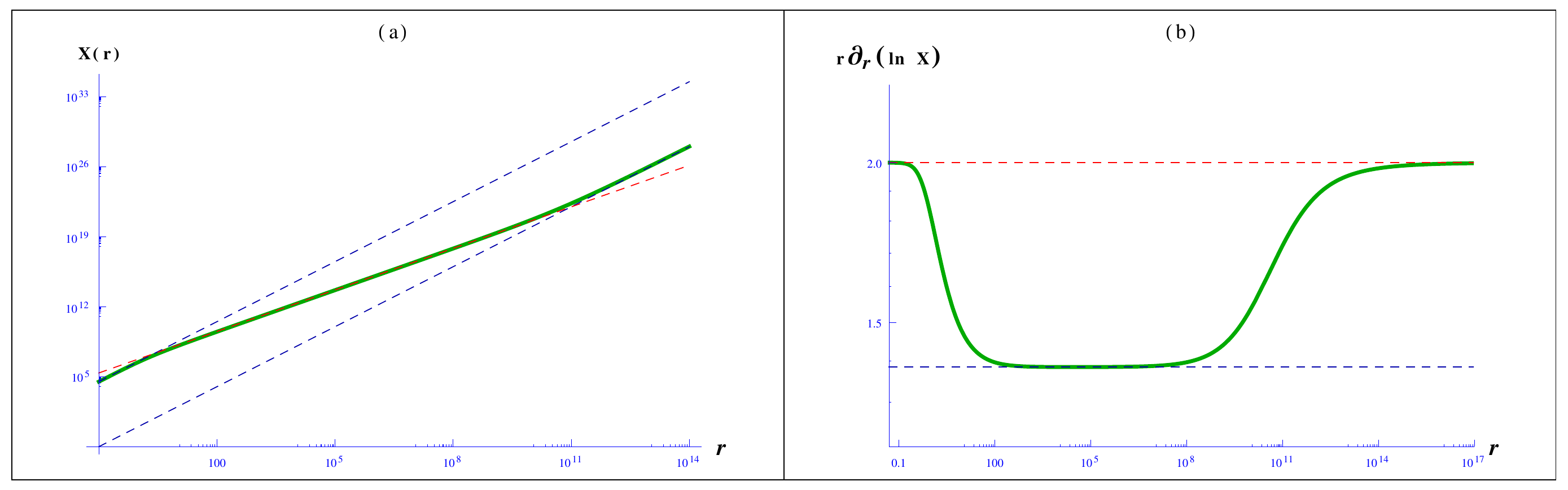}
\caption{\label{fig:A0z1Xplot} Numerical solution corresponding to the lagrangian parameters \eqref{A0z1para}. The solid green curves represent
log-log plots for the metric function $X(r)$ in (a) and its logarithmic derivative in (b).
The dashed straight lines confirm that $X(r)$ scales as $r^{1.38}$ in the intermediate region and as $r^2$ both in the UV and IR. }
\end{figure}

We choose the following numerical values of the lagrangian parameters,
\begin{equation}\label{A0z1para}
 \delta = \frac{2}{3}, ~V_0 = - \frac{1}{2}, ~V_1 = 0.5 \times 10^{-16}, ~V_3 = -5, ~Z_0 = 0, ~W_0 =0.
\end{equation}
Note that $V_1 \ll V_0, V_3$, implying a large separation between the two $AdS$ geometries, and a long intermediate $\theta \neq 0$ region.
The plot for the metric function is shown by the green curves in fig. \ref{fig:A0z1Xplot}.
For the set of parameters \eqref{A0z1para}, an intermediate regime of hyperscaling violation is expected with $\theta = -1.6$.
In turn this implies a scaling in the function $X \sim r^{1.38}$. In fig. \ref{fig:A0z1Xplot}(a) we provide
the log-log plot of $X(r)$; the slope of the red dashed line is precisely 1.38.
Thus, the numerical solution coincides with the analytical one in the intermediate regime.
In fig. \ref{fig:A0z1Xplot}(b) we show the log-log plot of the logarithmic derivative\footnote{
Notice that the logarithmic derivative plots are constructed so that any function scaling as $r^\alpha$ will appear as
a horizontal line with intercept at $\alpha$, making any scaling behavior readily apparent.
Thus, we shall use such plots in the rest of this section.
In fig. \ref{fig:A0z1Xplot} we present these plots alongside ordinary log-log plots of the function
to present a direct comparison.} of the same function $X(r)$.
Here the horizontal blue line is again drawn for the value 1.38, and its overlap with
the numerical solution confirms the regime of hyperscaling violation in the intermediate region.
Also note that in fig. \ref{fig:A0z1Xplot}(b) both in the IR and UV the solution asymptotes to
the red dashed line at 2, implying a scaling $\sim r^2$, which demonstrates the emergence
of $AdS_4$ geometries at both ends.

\begin{figure}[h!]
\centering
\includegraphics[scale=0.6]{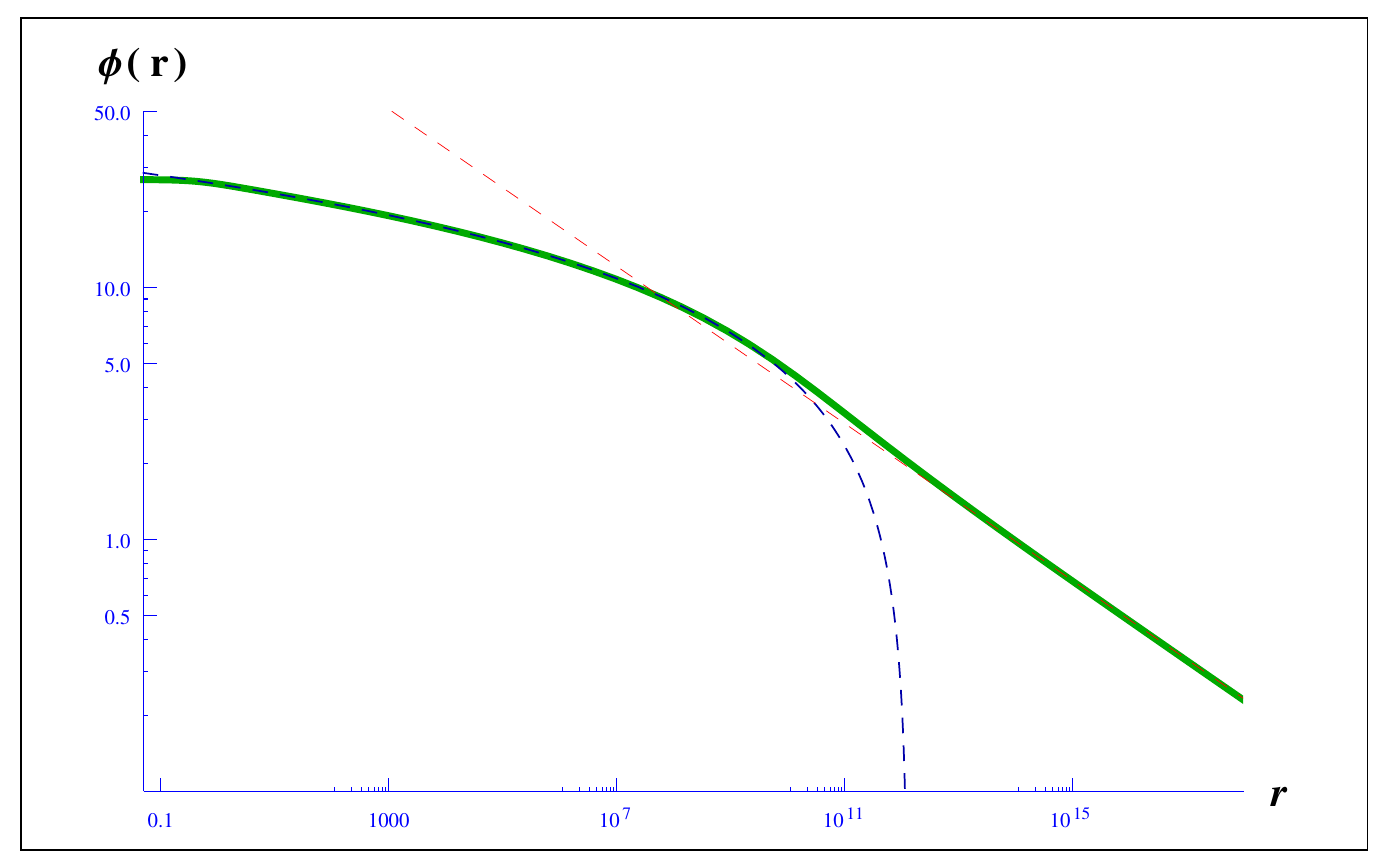}
\caption{\label{fig:A0z1phiplot}Log-log plot of the scalar field for the solution corresponding to the lagrangian parameters \eqref{A0z1para}.
The solid green curve represents the numerical solution. The dashed blue line shows intermediate logarithmic behavior while
the dashed red line shows that the dilaton goes to zero in the UV as $r^{-0.156}$, consistent with analytical
expectations.}
\end{figure}

The plot of the scalar field for this solution is shown in fig. \ref{fig:A0z1phiplot}. In the intermediate
regime it has a logarithmic behavior identical to \eqref{EMDScalingCohesivez=1} as is confirmed by the dashed blue line.
In the UV it goes to zero (its value at the maximum of the scalar potential) with an exponent $r^{-0.156}$. This is in accordance with the
expectation in equations \eqref{UVsclform}, \eqref{UVsd} and \eqref{UVmass}, and shows that we have a non-zero boundary source,
for the operator dual to dilaton.
Through a few numerical trials it seems that, once we have a sufficiently small $V_1$,
the presence of an intermediate hyperscaling violation regime is quite generic (irrespective
of the amplitude of the IR mode or equivalently the value of the scalar source in the UV).
Therefore, for this case it may be possible to fine tune the IR mode
to obtain a flow with vanishing boundary source. This is in contrast to the case
of $z \neq 1$, for which this may not be possible (see the discussion below).

Before closing this discussion, we would like to mention one curious fact about the solution \eqref{EMDScalingCohesivez=1}.
Naively this solution does not seem to exist when $\delta = -\eta = 1$.
However if we work in the gauge \eqref{newgauge} it can be shown that
a solution with the following geometry exists,
\begin{equation}
 ds^2 = \frac{1}{r} \, dr^2 - r \, dt^2 + r  \left( dx^2 + dy^2\right) \, ,
\end{equation}
the dilaton being logarithmic as usual. In some sense, this can be thought of
as a $\theta \rightarrow \infty$ limit of \eqref{EMDScalingCohesivez=1}; but the limit
needs to taken carefully, in a proper gauge. We can very easily incorporate this geometry in the intermediate region of the spacetime,
just like the one in fig. \ref{fig:A0z1Xplot} and fig. \ref{fig:A0z1phiplot}, which would make it perfectly regular.
We will not say anything more about this solution.

\subsubsection{Finite density flows with $z=1$ and \texorpdfstring{$\theta \neq 0$}{theta different from 0}}
\label{sssec:z=1}

When the gauge field does not backreact on the geometry it is also not possible to move away from the $z=1$ case.
However, in this case we can have a finite boundary chemical potential.
For this purpose, let us choose the following set of parameters
\begin{equation}\label{Aprobez1para}
 \delta = 0.9, ~V_0 = - 0.5, ~V_1 = 0.5 \times 10^{-20}, ~V_3 = -5, ~Z_0 = 1, ~W_0 =\frac{12}{23}, ~\alpha = - \frac{1}{3} \, .
\end{equation}
If we ensure that the dilaton is positive throughout the flow (which should be true since we are flowing
form the positive minimum to the maximum), then with $\alpha<0$
at leading order the terms $Z F^2$ and $W A^2$ can be neglected in the intermediate region when solving Einstein's equations
and the scalar equation of motion.
Therefore, for this set of parameters \eqref{Aprobez1para} the intermediate behavior is again expected to be
of the form \eqref{EMDScalingCohesivez=1}, with the gauge field scaling according to \eqref{probeAexpo}
(after the appropriate gauge change specified by \eqref{newgauge}).

\begin{figure}[h!]
\centering
\includegraphics[scale=0.56]{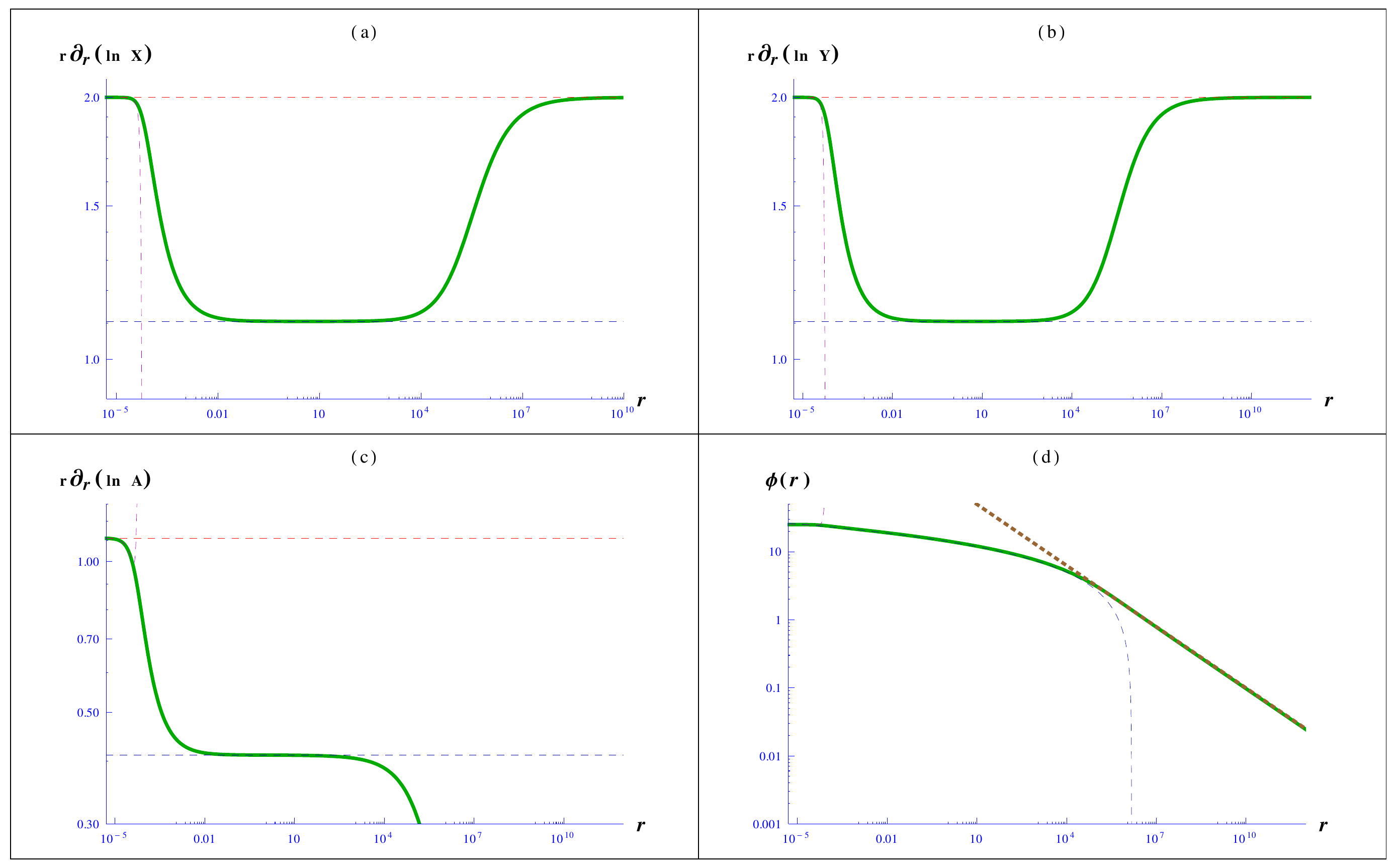}
\caption{\label{fig:Aprobez1}Numerical solution corresponding to the parameter set \eqref{Aprobez1para}. The solid green curves in (a), (b) and (c)
show the log-log plots of the logarithmic derivative of the metric functions and the gauge field respectively, and (d) shows the
log-log plot of the scalar field. The metric becomes $AdS_4$ both in the UV and in the IR, which is apparent from the $\sim r^2$ scalings
indicated by the dashed red lines.
In the intermediate region the metric functions scale as $\sim r^{1.10}$, while the gauge field like
$\sim r^{0.41}$ (both scalings are indicated by the dashed blue lines). The intermediate scaling exponents are $z=1$, $\theta=-8.5$.
The scalar field has the expected logarithmic behaviour in the intermediate region, as shown by the dashed blue line.
In the UV it goes to zero as $r^{-0.3}$ (shown by the dotted brown line), implying
the presence of a non-trivial boundary source, consistent with the expectation from \eqref{UVsclform}.
As a consistency check, we have plotted (with dashed magenta lines) the IR series solution to show
agreement with our numerical solution in a considerable IR region (although we have used it to set boundary conditions only at a single point).}
\end{figure}

The numerical solution for the parameter choice \eqref{Aprobez1para} is displayed in fig. \ref{fig:Aprobez1}, and has intermediate scaling exponents $z=1$, $\theta=-8.5$.
Note that the metric functions $X(r)$ and $Y(r)$ are identical, just like in the previous case, precisely because
the gauge field is not able to considerably backreact on the geometry.

%

\subsubsection{Finite density flows with \texorpdfstring{$z\neq1$}{z different from 1} and \texorpdfstring{$\theta \neq 0$}{theta different from 0}}
\label{sssec:zneq1}

\begin{figure}[h!]
\centering
\includegraphics[scale=0.56]{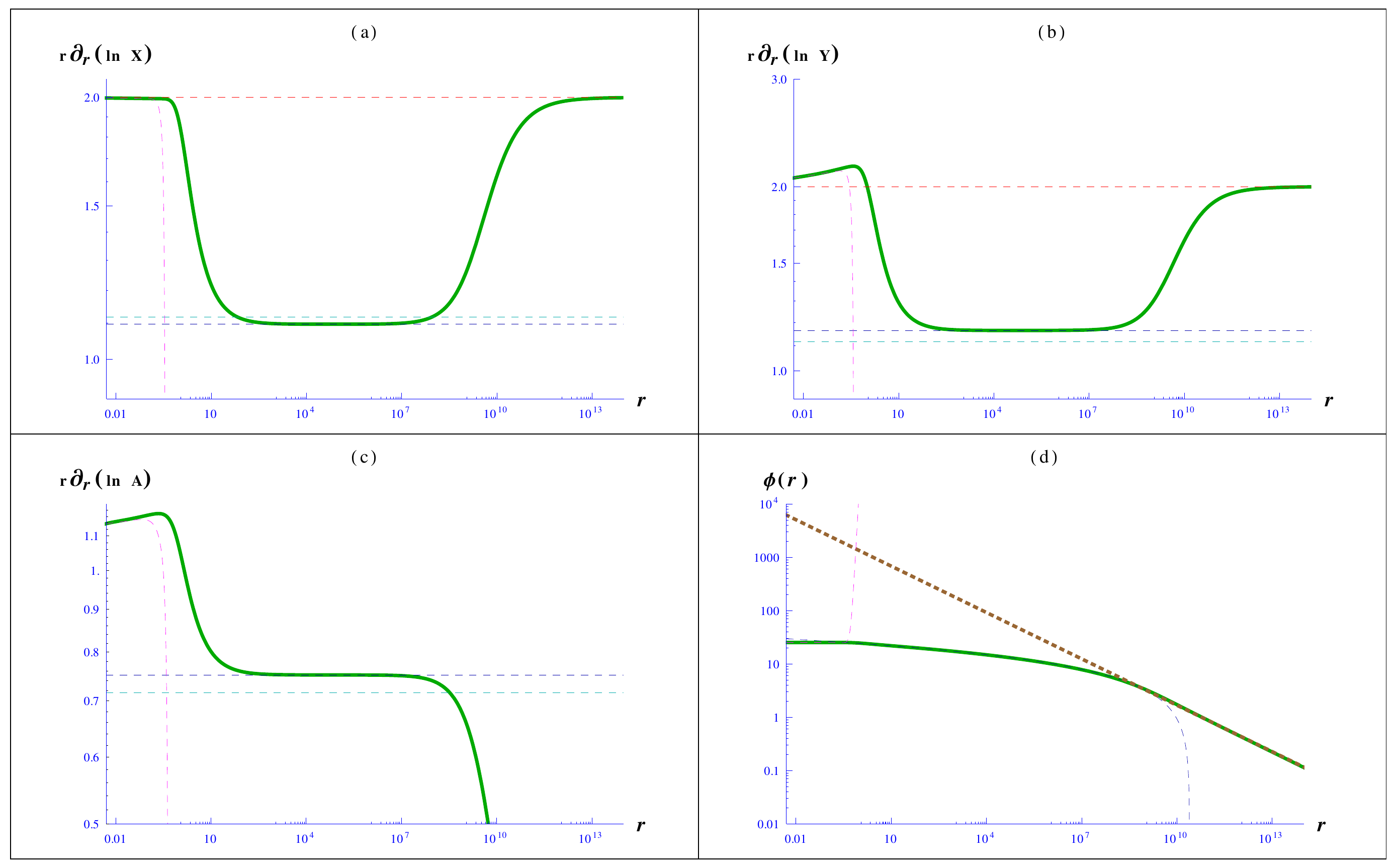}
\caption{\label{fig:zneq1}Numerical solution corresponding to the parameter set \eqref{zneq1para}. The solid green curves in (a), (b) and (c)
show the log-log plots of the logarithmic derivative of the metric functions and the gauge field respectively, and (d) shows the
log-log plot of the scalar field.
The metric becomes $AdS_4$ both in the UV and in the IR, as seen from the $r^2$ scalings
of $X(r)$ and $Y(r)$ indicated by the dashed red lines.
In the intermediate region the metric functions $X(r)$ and $Y(r)$ and the gauge field scale, respectively,
like $r^{1.09}$, $r^{1.16}$ and $r^{0.75}$ (indicated by the dashed blue lines). The intermediate scaling exponents are $z=1.3$, $\theta=-9.2$.
The scalar field has the expected logarithmic behaviour in the intermediate region (as
shown by the dashed blue line) and goes to zero as $r^{-0.29}$ in the UV (shown by the dotted brown line),
consistent with \eqref{UVsclform} and implying the presence of a non-trivial boundary source.
Again, we have plotted the IR series solution with dashed magenta lines, showing good agreement with the numerical solutions deep in the IR.
For comparison we have also included the $z=1$ scaling behavior of these functions (shown by the horizontal cyan lines),
which would describe the case of negligible backreaction of the gauge field on the geometry.}
\end{figure}

Finally we would like to obtain a hyperscaling violating solution with a non-trivial dynamical exponent ($z \neq 1$).
In order to achieve this we should ensure that the gauge field is of considerable strength. We now choose
the lagrangian parameters to be
\begin{equation}\label{zneq1para}
 \delta = \frac{8}{9}, ~V_0 = - 0.5, ~V_1 = 0.5 \times 10^{-20}, ~V_3 = -5, ~Z_0 = 1, ~W_0 =\frac{12}{23}, ~\alpha = \frac{1}{3} \, .
\end{equation}
Note the important change of sign of $\alpha$ compared to the previous case \eqref{Aprobez1para}. This ensures that the terms in the
lagrangian involving the gauge fields are comparable to the remaining terms, as long as the dilaton takes
positive values throughout the RG flow. As a result, in the intermediate region we obtain scaling solutions of the form \eqref{EMDScalingCohesive}.
The numerical plots for this case are displayed in fig. \ref{fig:zneq1}, and have $z=1.3$, $\theta=-9.2$.

 It should be pointed out that, unlike the $z=1$ cases discussed above, here the intermediate region matches with hyperscaling
 violating solution \eqref{EMDScalingCohesive} only after some fine tuning of the two available IR modes (corresponding to the gauge
 field and the scalar field). In fact, numerical scanning through various values of the amplitudes of the two IR modes
 seems to suggest that both of them are completely fixed once we fix the boundary chemical potential and demand that
 we have an intermediate hyperscaling violating Lifshitz solution.
 Therefore, we are left with no additional IR parameters
 which we can adjust to set the dilaton source at the boundary to zero\footnote{Note that, unlike for the $z=1$ case,
 here the amplitude of the gauge field in the intermediate region is not a free parameter.}.

\section{Conductivity}
\label{Conductivity}
%

Given that the domain wall solutions we have constructed exhibit an intermediate scaling region,
it is natural to ask if traces of this scaling can also be seen in the behavior of the optical conductivity.
First, note that in the deep IR the latter will already have power-law dependence on the frequency,
because of the emergent conformal symmetry of the underlying $AdS_4$ geometry \cite{Gubser:2008wz}.
At 
higher frequencies, however, we expect a transition to a region in which the conductivity will scale in a manner controlled
by the hyperscaling violating exponent $\theta$ and the dynamical critical exponent $z$ when $z\neq 1$,
and by the conduction exponent $\zeta$ when $z=1$ \cite{Gouteraux:2013oca}.
Finally, at the $AdS_4$ UV fixed point the conductivity should settle to a constant \cite{Herzog:2007ij}.

We would like to briefly remind the reader how to compute the optical conductivity $\sigma(\omega)$
and in particular how to rewrite it in terms of a Schr\"odinger problem \cite{Horowitz:2009ij}, for theories of the form
\beq
\label{repeatedLag}
\mathcal{L} = R - \half (\p\phi)^2 - \frac{Z(\phi)}{4} F^2 - \half W (\phi) A^2 - V(\phi) \, .
\eq
A similar analysis can be found in \cite{Goldstein:2009cv}, in which however there was
no mass term for the gauge field.
While for now we keep the scalar couplings and potential completely arbitrary,
recall that in the intermediate region they will have to reduce to
\beq
Z(\phi) = Z_0 e^{\alpha \phi}\, , \qquad  W(\phi) = W_0 e^{\beta \phi} \, , \qquad V(\phi) = V_0 e^{-\eta\phi} \, ,
\eq
in order to support the regime of hyperscaling violation and Lifshitz scaling.

The holographic recipe for extracting the conductivity entails solving for the appropriate linearized perturbation
of the gauge field. In particular, one perturbs both the background
gauge field (\ref{backgroundgg}) and the metric (\ref{metricansatz}) while leaving the scalar untouched,
\bea
ds^2 &=& - D(r) dt^2 + 2 \delta g_{tx} (t,r) dt dx + B(r) dr^2 + C(r) (dx^2 + dy^2) \, , \\
A &=& A_t(r) dt + \delta A_x (t, r) dx \, , \qquad \phi=\phi(r) \, ,
\ea
since $\delta g_{tx}$ and $\delta A_x$ contribute to the same order in perturbation theory.
The fluctuations can be chosen to have the simple form
\bea
\delta g_{tx}(t,r) &=& g_{tx}(r) e^{-i\omega t} \, , \\
\delta A_x(t,r) &=&  a_x(r) e^{-i\omega t} \, .
\ea
After appropriately massaging the perturbation equations,
$g_{tx}$ can be eliminated and the gauge field fluctuation $a_x(r)$ can be shown to obey
\beq
\label{axeom1}
\p_r \left( Z \sqrt{\frac{D}{B}}\, \p_r \, a_x \right) + \left[Z\sqrt{\frac{B}{D}} \, \omega^2
- \frac{Z^2}{\sqrt{BD}}(\p_r A_t)^2 - W \sqrt{BD}\, \right] a_x =0 \, .
\eq
Finally, after rescaling the gauge field fluctuation $a = \sqrt{Z} a_x$ and introducing a new radial coordinate\footnote{With our
parametrization for $AdS_4$, we have $\rho \sim - 1/r$ in the deep IR and far UV. Thus, in the new coordinate system
the horizon is at $\rho \rightarrow -\infty$ and the boundary at $\rho =0$.} $\rho$
through $d\rho/dr = \sqrt{B(r)/D(r)}$, the perturbation equation
(\ref{axeom1}) can be written in the form of a Schr\"odinger equation
\beq
\label{Seom}
- a^{\prime\prime}   + \mathcal{V} \, a = \omega^2 \, a \, ,
\eq
where $^\prime\equiv\p_\rho$ and the potential is given by
\beq
\label{Spotential}
\mathcal{V} = \frac{Z}{B} (\p_r A_t)^2 + \frac{W D}{Z} - \frac{Z^{\prime\; 2}}{4Z^2} +  \frac{Z^{\prime\prime}}{2Z}
= \frac{Z \, A_t^{\prime \; 2}}{D}  + \frac{W D}{Z} - \frac{Z^{\prime\; 2}}{4Z^2} +  \frac{Z^{\prime\prime}}{2Z} \, .
\eq
We see explicitly that the role of the gauge field mass term $\sim W$ is simply to contribute to the effective mass of the gauge field fluctuation.
The calculation of the optical conductivity is then essentially reduced to a Schr\"odinger problem.

\subsection{Matched Asymptotics}

The Schr\"odinger potential (\ref{Spotential}) can be easily evaluated in the near-horizon region of the geometry,
where it often takes the form
\beq
\label{V0}
\mathcal{V} (\rho) = \frac{\mathcal{V}_0}{\rho^2} \, ,
\eq
with the proportionality constant $\mathcal{V}_0$ determined by the particular form of the background.
The potential then vanishes at the horizon $\rho \rightarrow - \infty$ as discussed in \cite{Horowitz:2009ij}.
In particular, as we will see in more detail below, it is easy to verify (\ref{V0}) for our domain-wall geometries,
which approach $AdS_4$ in the deep IR.
The same behavior was also explicitly seen in \cite{Gouteraux:2013oca}
for the near-horizon region of $\{z,\theta\}$ scaling solutions\footnote{Where ensuring the heat capacity is still positive, $(2-\theta)/z$ means the IR is still $\rho\to-\infty$.}.

When the potential near the horizon is of the form (\ref{V0}), one can fix the
the small-$\omega$ scaling of the conductivity by using the method of matched asymptotics
developed in \cite{Gubser:2008wz}, and the fact that there is a conserved flux
$\p_\rho \mathcal{F} = 0$ associated with the Schr\"odinger equation (\ref{Seom}).
Here we would like to briefly review the main steps behind the steps of \cite{Gubser:2008wz},
later elaborated on in \cite{Horowitz:2009ij}.
First, the real part of the conductivity can be extracted from the UV value of the flux,
\beq
\mathcal{F} \sim \omega \;  |a^{(0)}|^2 \; \text{Re}(\sigma) \, ,
\eq
once we have determined $a^{(0)}$, the leading component of the gauge field fluctuation at the boundary.
In our construction the conserved flux is given by
\beq
\label{Flux}
\mathcal{F}= i a^\ast \overleftrightarrow{\p_\rho} a =
i \sqrt{\frac{D}{B}} \left[ \left(\sqrt{Z} a_x\right)^\ast  \p_r \left(\sqrt{Z} a_x\right)- \sqrt{Z} a_x \p_r
\left(\sqrt{Z} a_x\right)^\ast \right] \, .
\eq
Precisely because it is a conserved quantity, the flux can equally well be computed near the extremal horizon.
To do so one needs to have the solutions to the Schr\"odinger equation in the deep IR, where the potential is well approximated by (\ref{V0}).
It is easy to show that in this case the near-horizon solutions are Hankel functions,
with the purely ingoing\footnote{As $\rho \rightarrow - \infty$
the solution becomes
$ a \sim e^{- i \omega \rho}$.} mode at the horizon given by
\beq
\label{index}
a \sim \sqrt{\omega \rho} \; H_\nu^{(1)}(\omega \rho) \, , \qquad \nu = \sqrt{\mathcal{V}_0 + \frac{1}{4}} \, .
\eq
Thus, in the near-horizon region the flux is found by evaluating (\ref{Flux}) on (\ref{index}).

The last ingredient needed to obtain $\sigma(\omega)$ is the boundary behavior of the fluctuation, and specifically
the way in which $a^{(0)}$ scales with frequency.
To solve for the boundary fluctuation, one can use the method of matched asymptotics.
The logic behind it is that -- for $\omega$ sufficiently small --
there exists a window in which appropriate IR and UV solutions overlap, and can therefore be matched to each other.
It is then straightforward to extract the $\omega$-dependence of the gauge field fluctuation at the boundary.
Combining all these ingredients finally gives the scaling of the conductivity
\beq
\label{sigmascaling}
Re(\sigma) \sim \omega^{2\nu-1} \, , \qquad T \ll \omega \ll \mu \; \, ,
\eq
with the precise power controlled by the expression for $\mathcal{V}_0$ via (\ref{index}).
\vspace{0.2in}

\noindent
\emph{Scaling associated with the deep IR $AdS_4$ background}
\newline
The matching procedure we have just outlined can be implemented to extract the scaling of the conductivity
in the deep IR of our model, where it is sensitive to the $AdS_4$ background geometry.
First of all, recalling the IR behavior of the metric, scalar and gauge fields,
given respectively by (\ref{IRexp}), (\ref{irrscalar}) and (\ref{ggpert}),
it is easy to check that the only contribution to the Schr\"odinger potential (\ref{Spotential})
in the near-horizon region comes from the gauge field mass term, i.e.
$ \mathcal{V} \sim \frac{W(\phi_0)}{Z(\phi_0)}\, D$.
Since the metric function is given by $D (r) \sim r^2$ and near the horizon $\rho \sim - L_{IR}/r$,
it is immediately apparent that (\ref{V0}) holds, with the identification
\beq
\mathcal{V}_0 = \frac{L_{IR}^2 \, W(\phi_0)}{Z(\phi_0)}  \qquad \Rightarrow \qquad \nu = \half \, \sqrt{1+\frac{4 L_{IR}^2 W(\phi_0)}{Z(\phi_0)} } \, .
\eq
Thus, in the IR the mode which is purely ingoing at the horizon is given by (\ref{index}),
and will propagate in the bulk without being disturbed up to some scale $\sim \rho_1$.
Near the boundary, the Schr\"odinger potential diverges (see appendix \ref{app:SchrPot})
and the $\omega$ term in the Schr\"odinger equation can be neglected.
The equation can then be solved (e.g. in a small $\omega$ approximation),
and its near-boundary solution will still be valid at non-zero frequency and into the bulk
down to some scale $\rho_2 \sim -1/\omega_2$,
for some $\omega\ll\omega_2$.
If $\omega_2$ is small enough, then there is a large overlap window where both IR and UV solutions are valid,
$ \rho_2 \ll \rho\ll \rho_1<0$,
and where the matching can be done.
Finally, using (\ref{sigmascaling}) we expect the optical conductivity in the deep IR to scale as
\beq
\label{sigmaIR}
Re(\sigma) \sim \omega^{2(\Delta_J-2)} \, ,
\eq
with a power dictated by the conformal dimension (\ref{DeltaJ}) of the current, as in \cite{Gubser:2008wz}.
\vspace{0.2in}

\noindent
\emph{Scaling in the Lifshitz and hyperscaling violating regime}
\newline
The same procedure described above applies to $\{z,\theta\}$ scaling solutions to (\ref{repeatedLag}) which extend
all the way into the deep IR of the geometry \cite{Gouteraux:2013oca}.
Although this is not the case in our setup,
we briefly state the results of the analysis of \cite{Gouteraux:2013oca}, as they will be relevant for our discussion:
\begin{itemize}
\item
when the IR geometry is the cohesive $z\neq 1$ hyperscaling violating solution (\ref{EMDScalingCohesive})
(corresponding to $\beta = \alpha - \eta$), one finds that (\ref{V0}) holds  and
\beq
\mathcal{V}_0 = \frac{ (2 z - \theta)(4z - \theta)}{4z^2} \, .
\eq
In turn the conductivity scales as
\beq
\label{sigmaznot1}
Re(\sigma) \sim \omega^{2-\frac{\theta}{z}} \, .
\eq
\item
when the IR geometry is the $z=1$ hyperscaling violating solutions (\ref{EMDScalingCohesivez=1}) corresponding to
a perturbatively small gauge field $A_t^{IR} \sim r^{2\frac{\zeta-\xi-1}{\theta-2}}$ (whose electric flux scales as $\sim r^{\frac{2\xi}{\theta-2}}$),
again (\ref{V0}) holds, with the identification
\beq
\mathcal{V}_0 = \frac{\zeta(\zeta-2)}{4} \, ,
\eq
implying a scaling for the conductivity given by
\beq
\label{sigmaz1}
Re (\sigma) \sim \omega^{-\zeta} \,,
\eq
hence the name conduction exponent for $\zeta$. Note also that this is the scaling which would be observed for zero density solutions where $A_t=0$. Upon setting $\zeta=\theta-2$, this agrees with the $z=1$ limit of \eqref{sigmaznot1}.
\end{itemize}
Since in our construction the $\{z,\theta\}$ scaling behavior appears only
in an intermediate regime, in order to probe it one needs to reach higher values of the frequency.
Given that the matched asymptotics analysis relies crucially on having small frequencies, it is not a priori
obvious that the $\{z,\theta\}$ scalings would be visible in the optical conductivity, in an appropriate
frequency regime.
As will be clear from the numerics, however, $\sigma(\omega)$ does indeed exhibit two distinct scaling regimes\footnote{With a caveat for zero density solutions, see the discussion further below in section \ref{ssec:numcond}.}
-- one in the deep IR due to the emergent conformal symmetry, and one
at higher frequencies due to hyperscaling violation --
with the precise powers in agreement with the predictions of the
standard matched asymptotics analyses (\ref{sigmaznot1}) and (\ref{sigmaz1}).
To explain why the agreement with (\ref{sigmaznot1}) and (\ref{sigmaz1}) works so well, we note that
if the matching between the UV and the IR solutions is done high up in the
intermediate scaling region (and not in the $AdS_4$ region),
then effectively the incoming near-horizon mode will be that of the hyperscaling violating region\footnote{We won't
have excited any of the outgoing modes of the intermediate scaling geometry.}.
For an appropriate frequency range, the intermediate solution will overlap with the near-boundary one, so they can be matched.
Thus, in this case the deep IR $AdS_4$ geometry plays no role in the matching procedure, and does not disturb the logic
behind (\ref{sigmaznot1}) and (\ref{sigmaz1}).

\subsection{Numerical plots of conductivity} \label{ssec:numcond}
%

\begin{figure}[h!]
\centering
\includegraphics[scale=0.56]{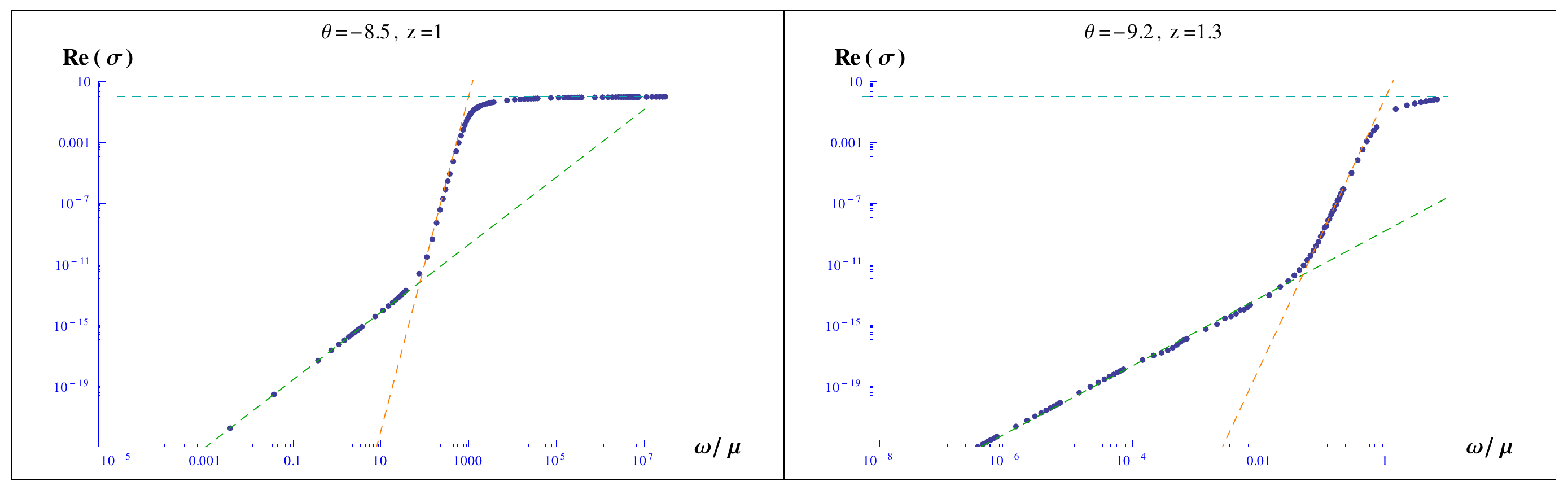}
\caption{\label{fig:conductivity} Log-log plots of optical
conductivity vs. frequency (normalized by the chemical potential),
corresponding to the backgrounds in fig. \ref{fig:Aprobez1} (left panel) and fig. \ref{fig:zneq1} (right panel).
The dashed green lines display the IR scaling determined by the emergent IR conformal fixed point, which for these two cases is
a scaling of $\omega^{2.22}$ (consistent with \eqref{sigmaIR} and \cite{Gubser:2008wz}).
The orange lines represent the scaling due to the intermediate regime of hyperscaling violation.
The left panel displays a scaling of $\omega^{10.99}$ (consistent with \eqref{sigmaz1}),
while the right panel shows a scaling of $\omega^{8.93}$ (in agreement with \eqref{sigmaznot1}).
In the UV the conductivity saturates to unity in both plots, as denoted by the cyan line.}
\end{figure}

Following the discussion above, it is straightforward to study numerically the frequency dependence of the
optical conductivity at zero temperature.
We will do so first for the \emph{finite density} $z=1$ and $z\neq 1$ backgrounds shown, respectively,
in fig. \ref{fig:Aprobez1} and fig. \ref{fig:zneq1}.
The numerical plots of the optical conductivity for these two cases are shown in fig. \ref{fig:conductivity}.
Note the existence of three distinct regimes.
In both panels the low-frequency behavior of $\text{Re}(\sigma (\omega))$ agrees with the analytical expectation (\ref{sigmaIR})
for the emergent conformal fixed point, and is fixed by the conformal dimension of the current in the IR.
In the middle region, on the other hand, the scaling is that predicted by (\ref{sigmaz1}) for the $z=1$ background
(in the left panel), and by (\ref{sigmaznot1}) for the $z\neq 1$ case (in the right panel).
Thus, we see a clean intermediate scaling regime controlled by the exponents of the underlying geometry.
Finally, for sufficiently high frequencies only the UV $AdS_4$ is probed, where the conductivity is unity -- the standard result
expected for the normalization we have chosen.

We can estimate the scales at which these transitions happen using the radial scales determined by inspection of our numerical flows in section \ref{ssec:numsol}. 
They will occur when $\omega\rho\sim1$ (which is where the approximate solution of the fluctuation in terms of Hankel function ceases to be valid), 
where $\rho$ is the Schr\"odinger coordinate defined below \eqref{axeom1}, approximating the metric functions by 
their expressions in the IR AdS$_4$ or intermediate hyperscaling violating regime. The estimates agree roughly with the numerical results.

At this point it is important to emphasize that in the constructions above the scalings in the intermediate regime are \emph{always positive},
as apparent from the numerical plots.
This is a consequence of imposing a number of consistency conditions -- ensuring that the
Null Energy Condition holds, that the specific heat is positive and the electric flux is monotonic -- as well as
requiring the existence of appropriate irrelevant deformations about the intermediate geometries \cite{Gouteraux:2013oca}.

\begin{figure}[h!]
\centering
\includegraphics[scale=0.56]{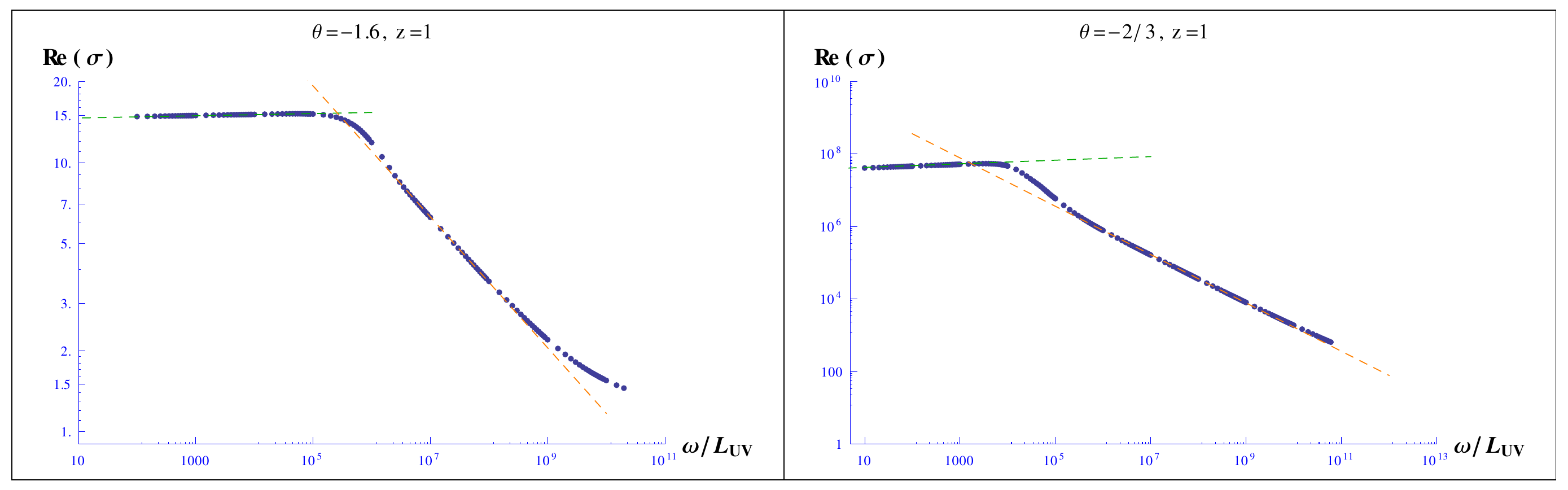}
\caption{\label{fig:conductivityneg} Conductivity plots showing an intermediate scaling regime characterized by a negative
exponent. In the left panel we have used the background shown in fig.\ref{fig:A0z1Xplot} and fig.\ref{fig:A0z1phiplot},
with the parameters of \eqref{negcondapara}. The orange line represents
a scaling of $\omega^{-0.243}$ according to \eqref{sigmaz1}. In the right panel the parameter choice is that of
\eqref{negcondbpara}. Here the intermediate scaling (the orange line) is given by $\omega^{-2/3}$.
In both plots the dashed green lines display the IR scaling as in \eqref{sigmaIR},
determined by the emergent IR conformal fixed point in accordance with \cite{Gubser:2008wz}.}
\end{figure}

Some of these conditions (and in particular the structure of vector deformations about the intermediate region)
can, however, be relaxed by working at \emph{zero density}. Negative intermediate scalings can then be engineered, as was previously highlighted in \cite{Charmousis:2010zz} for zero vector mass.
In fig. \ref{fig:conductivityneg} we include two numerical plots of the conductivity in the zero
density case, where we exhibit an intermediate scaling regime with a negative exponent\footnote{Positive scalings, similar to the ones previously reported on at finite density, can also
be observed.}.
The latter is controlled by the conduction exponent $\zeta$ as in \eqref{sigmaz1}.
For the plot in the left panel of fig.\ref{fig:conductivityneg} we use the background solutions presented in fig.\ref{fig:A0z1Xplot}
and fig.\ref{fig:A0z1phiplot}, with the parameter choice
\begin{equation}\label{negcondapara}
 \delta = \frac{2}{3}, ~V_0 = - \frac{1}{2}, ~V_1 = 0.5 \times 10^{-16}, ~V_3 = -5, ~Z_0 = 1, ~W_0 = 4.57 \times 10^{-4}, ~\alpha = 0.105.
\end{equation}
For this case, we obtain an intermediate scaling which goes as $\omega^{-0.243}$ (according to \eqref{sigmaz1}),
as denoted by the orange dashed line.
In the right panel we have chosen instead the following set of parameters
\begin{equation}\label{negcondbpara}
  \delta = \frac{1}{2}, ~V_0 = - 0.5, ~V_1 = 0.5 \times 10^{-14}, ~V_3 = -5, ~Z_0 = 1, ~W_0 = 5.68 \times 10^{-3}, ~\alpha = \frac{3}{4} \, ,
\end{equation}
for which we can obtain an intermediate scaling of the form $\omega^{-2/3}$.
In both cases, the IR scaling behavior is precisely as predicted by \eqref{sigmaIR} (see \cite{Gubser:2008wz}).
Note that it is technically time consuming to generate points for higher frequencies,
while keeping the numerical precision in the numerical evaluation intact.
Consequently, the UV behavior is not manifest in the plots in fig.\ref{fig:conductivityneg}.
However, we have verified explicitly that the Schr\"odinger potential \eqref{Spotential} has the appropriate UV scaling properties
which guarantee that the conductivity at higher frequencies should settle down to unity, just as in
fig.\ref{fig:conductivity}.

Importantly, we close this section with a subtlety which we noted when the conductivity scaling is negative in the intermediate regime, at zero density. For some region of the parameter space, we did not observe an intermediate scaling regime, and the appearance of a bump. We could correlate this to the appearance of a negative well in the Schr\"odinger potential, at the interplay between the IR $AdS$ regime and intermediate hyperscaling violating regime. Upon suppression of this well, the intermediate scaling regimes displayed in fig. \ref{fig:conductivityneg} were recovered. For deep enough wells, a bound state may form, which signals an instability of the neutral CFT plasma and is accompanied by the crossover of a pole from the lower half-plane to the upper half-plane. While it is not clear that this will always spoil the intermediate scaling, the conductivity calculation can no longer be trusted because of the instability.

The first appearance of a bound state at zero energy in a potential well with no vertical walls can be more precisely estimated by using the formula
\beq
\left(n-\frac12\right)\pi=\int_{\rho_1}^{\rho_2}\sqrt{-V(\rho)}\,d\rho
\eq
where $n\geq1$ is an integer, derived using the WKB approximation in the Schr\"odinger equation. We will leave the study of this important feature for future work.

\section{Outlook}
\label{Outlook}

In this work, we have shown how hyperscaling violating geometries can be realised along holographic RG flows as intermediate
regimes between two bona fide $AdS_4$ fixed points in the IR and in the UV\footnote{Though we have been working in four bulk dimensions throughout, it is clear that our construction will carry through in more dimensions.}.
The putative tidal or naked curvature singularities of these solutions are regularized by the emergence of the IR $AdS_4$ scaling.
The crucial ingredient to achieve the IR $AdS_4$ is the presence of a massive vector field,
which breaks gauge invariance.
A natural way to motivate the latter is to view our model (\ref{LagMassive1})
 -- assuming mild conditions on the couplings -- as describing the
broken symmetry phase of a $U(1)$ theory coupled to a complex charged scalar.
In analogy with the standard holographic superconductor setup,
the condensation of the charged scalar can be responsible for the spontaneous breaking of the abelian gauge symmetry.
Although we haven't made this connection more explicit, we don't foresee any obstacles to doing so
for appropriate values of the lagrangian parameters.
To be consistent with that picture, we have chosen the effective mass $W(\phi)$ of the gauge field to vanish in the UV and
to be well approximated by $\sim \phi^2$ near the boundary.
We expect the main features of our construction -- most notably the intermediate scaling regime --
to survive in the cases in which the model can be realized
through spontaneous symmetry breaking. We have already noted however that obtaining flows with intermediate $z\neq1$ seems to require some fine-tuning of the boundary data, which may turn out to be incompatible in some cases with spontaneous symmetry breaking. This could place some interesting constraints on allowed RG flows.

We have shown zero-temperature flows which exhibit on top of hyperscaling violation either Lifshitz scaling in the intermediate regime,
or relativistic scaling throughout the flow. An important generalization is to construct finite temperature
black holes whose ground state would be the zero-temperature flows we have studied.
We then expect the different scaling regimes to be also displayed in the Hawking-Bekenstein entropy dependence on temperature.
We leave this for future work.

Importantly, this succession of scaling regimes is communicated to the zero temperature, frequency-dependent optical conductivity.
This is reminiscent of the phenomenology of certain condensed matter systems \cite{vandermarel1, vandermarel2}.
However, an important feature of these scalings in our construction is that they come with a positive power,
i.e. $\sigma (\omega) \sim \omega^n$, $n>0$, when the system is at finite density.
This is an inevitable consequence of imposing conditions such as the Null Energy Condition, positivity of the specific heat,
monotonicity of the electric flux\footnote{Though this particular condition could be relaxed in principle,
it is hard to imagine charges of opposite signs coexisting in the bulk and not annihilating each other.
See \cite{Nitti} however for oppositely charged shells in the context of electron stars.} and consistency of the deformations
around the intermediate geometries.
A first possible way around these limitations is to work at zero density,
since in this case some of these conditions (such as the structure of vector deformations about
the hyperscaling violating region) no longer apply. As a result, negative scalings for the conductivity, $\sigma (\omega) \sim \omega^n$, $n<0$, can be engineered at intermediate
frequencies, as we have shown in our numerics.

It would also be interesting to break translational symmetry, in order to get closer to real-life systems.
Two simple options retaining homogeneity of the field equations would be the introduction of massless,
spatially dependent scalars as in \cite{Donos:2013eha,AW,Donos:2014uba,g2014}, or of Bianchi VII lattice deformations \cite{Donos:2012js,Donos:2014oha}.
This would regulate the delta function present at finite temperature when translation symmetry is unbroken, and moreover $n<0$ could occur in conducting states.

Another important aspect we have encountered is the spoiling of the intermediate conductivity scaling regime, when a deep enough negative well develops in the Schr\"odinger potential (which still displays three well-defined scaling regimes itself). This happens at zero density for negative intermediate scalings. We interpret this as the formation of a bound state. Bound states are associated with an instability of the neutral CFT plasma, rendering the previous calculation of the conductivity unreliable. It would be interesting to understand better the parameter space where such bound states are disallowed, and whether they could also appear at finite density.

We also expect that the same intermediate scaling behaviour will be displayed in other transport observables such as the
heat or thermoelectric conductivities, as well as in their DC counterparts.

\acknowledgments

We would like to thank Jorge E. Santos for initial collaboration and valuable input.
We are also grateful to Pallab Basu, Shamik Banerjee, Aristomenis Donos,
Sean Hartnoll, Elias Kiritsis, Nilay Kundu, Charles Milton Melby-Thompson, Rene Meyer,
Subir Sachdev, Kai Sun and Tadashi Takayanagi for many useful conversations.
We would like to especially thank Ben Withers for many helpful comments
and suggestions on the numerical analysis, as well as Aristomenis Donos, Elias Kiritsis and Tadashi Takayanagi for valuable comments on the manuscript. The work of SC has been supported by
the Cambridge-Mitchell Collaboration in Theoretical Cosmology, and the Mitchell Family Foundation.
The work of BG was supported by the Marie Curie International Outgoing
Fellowship nr 624054 within the 7$^\textrm{th}$ European Community Framework Programme FP7/2007-2013.
The work of JB was supported by World Premier International Research Center Initiative (WPI Initiative), MEXT, Japan.

\appendix

\section{Schr\"odinger Potential Near the Boundary \label{app:SchrPot}}

We would like to examine the behavior of the Schr\"odinger potential (\ref{Spotential}) in our model as we approach the boundary $r \to \infty$,
and in particular its dependence on the conformal dimension of the scalar.
Recall that near the boundary $\phi = \phi_{UV} + \delta \phi$, where for us
$\phi_{UV}=0$, and in general
\beq
\delta \phi = \phi_a \, r^{-\Delta_+} + \phi_b \, r^{-\Delta_-}  =  \phi_a \, r^{-\Delta} + \phi_b \, r^{\Delta-3}  \, ,
\eq
where
\beq
\Delta_\pm = \frac{3}{2} \pm \sqrt {\frac{9}{4} + m_{UV}^2 L_{UV}^2 } \, .
\eq
In the standard quantization $\Delta = \Delta_+$ and $\phi_a$ denotes the VEV of the operator dual to the scalar, while
$\phi_b$ is the source (in the alternative quantization instead $\Delta = \Delta_-$ and the roles of the VEV and source are reversed).
For convenience we also recall that in $AdS_4$  the unitarity bound for a conformal primary operator
constrains the conformal dimension to be $\Delta \geq \half$.

We will estimate $\mathcal{V}$ first by keeping both source and VEV turned on, and then examine what happens when the source is
set to zero ($\phi_b=0$ using the standard quantization).
Recalling that in the UV $AdS_4$ the metric, gauge field and scalar behave as
\beq
D(r) \sim r^{2} \; ,  \qquad A_t \sim \mu + \frac{\rho}{r} \, , \qquad \phi \sim \delta \phi = \phi_a \, r^{-\Delta} + \phi_b \, r^{\Delta-3}  \, ,
\eq
we can easily estimate the boundary behavior of the Schr\"odinger potential,
\bea
\label{Vgeneral}
\mathcal{V} &=& c_1 r^{-2} +  \phi_a^2 \left(c_2 + \Delta^2(c_3 +c_4) \right) r^{2(1-\Delta)} +
 \phi_a \, c_5 \, \Delta(\Delta-1) \, r^{2-\Delta}  + \nn \\
&+&  \phi_b^2 \, \left[c_2 + (3-\Delta)^2(c_3 +c_4) \right] r^{2(\Delta-2)} +
 \phi_b \, c_5 \, (3-\Delta)(2-\Delta) \, r^{\Delta-1} \nn \\
 &+& 2 \phi_a \, \phi_b \left[ c_2 + (c_3 + c_4) \Delta (3-\Delta) \right] r^{-1} \, ,
\ea
where the $c_i$'s are constants. More precisely, $c_1 \propto \rho^2$ is associated with the flux, $c_2 \propto W_0$ with
the gauge field mass term, and the remaining coefficients come from the direct coupling between the scalar and the
gauge field kinetic term,
$c_3 \propto (Z^{\prime} (\phi_{UV}))^2$ and  $c_4, c_5 \propto Z^{\prime\prime} (\phi_{UV})$.

When both source and VEV are turned on, the potential (\ref{Vgeneral}) is always divergent
as $r\rightarrow \infty$ in the UV ($\mathcal{V} \rightarrow \pm \infty$ depending on the signs of the couplings),
except for the two cases with $\Delta = 1,2$, for which it becomes constant.
On the other hand when the source is zero, $\phi_b = 0$, (\ref{Vgeneral}) takes the simpler form
\beq
\label{Vnosource}
\boxed{
\mathcal{V} \sim c_1 \,  r^{-2} + \phi_a^2\, ( c_2 + \Delta^2 (c_3+c_4)) \,  r^{2(1-\Delta)}
+ c_5 \phi_a \, \Delta (\Delta-1) r^{2-\Delta}\, , \;}
\eq
which vanishes in the far UV when $\Delta >2$, is constant when $\Delta =2$ and $\Delta =1 $ and diverges
everywhere else (when $1<\Delta < 2 $ and $1/2 < \Delta <1$ if we were to impose the unitarity bound).
Finally, note that when the gauge kinetic coupling is independent of the scalar field
so that $c_3=c_4=c_5=0$, the expression (\ref{Vnosource}) reduces to that already seen in \cite{Horowitz:2009ij},
\beq
\mathcal{V} \sim c_1 \,  r^{-2} + c_2 \, \phi_a^2 \, r^{2(1-\Delta)} \, .
\eq
This expression vanishes when $\Delta>1$, it is constant when $\Delta=1$ and is divergent when $\half < \Delta <1$,
where we have implicitly imposed the unitarity bound on the conformal dimension.

\bibliographystyle{JHEP}
\bibliography{AdSHVAdS}

\end{document}